\documentclass[12pt]{article}%
\usepackage{amssymb}
\usepackage{graphicx}
\usepackage{amsmath}
\usepackage{amsfonts}
\usepackage{float}
\usepackage{color}
\usepackage{colortbl}%

\usepackage{tikz}
\usepackage{pgfplots}
\usepackage{calc}
\usetikzlibrary{calc}
\usetikzlibrary{positioning}
\usetikzlibrary{decorations.pathmorphing}
\usetikzlibrary{decorations.text}
\usetikzlibrary{backgrounds}
\tikzstyle{whitevertex}=[circle, fill=white, draw=black, thick]
\tikzstyle{blackvertex}=[circle, fill, draw=black, thick]

\usepackage{dcolumn}
\usepackage{longtable}
\usepackage{booktabs}
\usepackage{threeparttable}

\usepackage[labelformat=simple]{subcaption} 
	
\usepackage{caption, subcaption}
\usepackage[hidelinks]{hyperref} 

\newcolumntype{M}{>{\centering\arraybackslash}m{\dimexpr.25\linewidth-2\tabcolsep}}

\providecommand{\U}[1]{\protect\rule{.1in}{.1in}}
\newtheorem{theorem}{Theorem}[section]

\newtheorem{defn}[theorem]{Definition}

\newtheorem{lemma}[theorem]{Lemma}
\newtheorem{propn}[theorem]{Proposition}

\newtheorem{question}[theorem]{Question}
\begin{document}

\title{\textbf{The~$k$-conversion number of regular graphs}}
\author{C.~M.~Mynhardt\thanks{Supported by the Natural Sciences and Engineering
Research Council of Canada.} {\ } and J.~L.~Wodlinger\footnotemark[1]\\Department of Mathematics and Statistics\\University of Victoria, P. O. Box 3045, Victoria, BC\\\textsc{Canada} V8W 3P4\\{\small kieka@uvic.ca, jw@uvic.ca}}
\maketitle

\begin{abstract}
Given a graph~$G=(V,E)$ and a set~$S_{0}\subseteq V$, an irreversible~$k$%
-threshold conversion process on~$G$ is an iterative process wherein, for
each~$t=1,2,\dots$, ~$S_{t}$ is obtained from~$S_{t-1}$ by adjoining all
vertices that have at least~$k$ neighbours in~$S_{t-1}$. We call the
set~$S_{0}$ the seed set of the process, and refer to~$S_{0}$ as an
irreversible~$k$-threshold conversion set, or a~$k$-conversion set, of~$G$
if~$S_{t}=V(G)$ for some~$t\geq0$. The~$k$-conversion number~$c_{k}(G)$ is the
size of a minimum~$k$-conversion set of~$G$.

A set~$X\subseteq V$ is a decycling set, or feedback vertex set, if and only
if~$G[V-X]$ is acyclic. It is known that~$k$-conversion sets in~$(k+1)$%
-regular graphs coincide with decycling sets.

We characterize~$k$-regular graphs having a~$k$-conversion set of size~$k$,
discuss properties of~$(k+1)$-regular graphs having a~$k$-conversion set of
size~$k$, and obtain a lower bound for~$c_{k}(G)$ for~$(k+r)$-regular graphs.
We present classes of cubic graphs that attain the bound for~$c_{2}(G)$, and
others that exceed it---for example, we construct classes of~$3$-connected
cubic graphs~$H_{m}$ of arbitrary girth that exceed the lower bound for
$c_{2}(H_{m})$ by at least~$m$. 

\end{abstract}

\noindent\textbf{Keywords:\hspace{0.1in}}irreversible~$k$-threshold conversion
process,~$k$-conversion number, decycling set, decycling number, cubic graph

\noindent\textbf{AMS Subject Classification Number 2010:\hspace{0.1in}}05C99,
05C70, 94C15%

\definecolor{medlightgray}{RGB}{206,206,206}%
\definecolor{medgray}{RGB}{170,170,170}%
\definecolor{darkgray}{RGB}{128,128,128}%
\thispagestyle{empty}

\section{Introduction}

Given a graph~$G=(V,E)$ and a set~$S_{0}\subseteq V$, an \emph{irreversible~}%
$k$\emph{-threshold conversion process} on~$G$ is an iterative process
wherein, for each~$t=1,2,\dots$, ~$S_{t}$ is obtained from~$S_{t-1}$ by
adjoining all vertices that have at least~$k$ neighbours in~$S_{t-1}$. We call
the set~$S_{0}$ the \emph{seed set} of the process, and refer to~$S_{0}$ as an
\emph{irreversible~}$k$\emph{-threshold conversion set}, or simply a
$k$\emph{-conversion set}, of~$G$ if~$S_{t}=V(G)$ for some~$t\geq0$. The
$k$\emph{-conversion number}~$c_{k}(G)$ is the size of a minimum
$k$-conversion set of~$G$.

A set~$X\subseteq V$ is a \emph{decycling set}, or \emph{feedback vertex set},
if and only if~$G[V-X]$ is acyclic. Early research on decycling sets was
motivated by applications in logic networks and circuit theory, first in
digraphs~\cite{DivietiGrasselli1968, LempelCederbaum1966} and later in
undirected graphs~\cite{Harary1975}. More modern applications are given
in~\cite{Honma+2016}. The \emph{decycling number}~$\phi(G)$ of a graph~$G$ is
the size of a minimum decycling set of~$G$. Clearly, finding a minimum
decycling set of~$G$ is equivalent to finding a maximum induced forest. The
order of such a forest is called the \emph{forest number} of~$G$, and denoted
by~$a(G)$. Many authors have derived bounds on~$\phi(G)$ and~$a(G)$, both for
general graphs~\cite{BeinekeVandell1997} and for special classes of graphs,
including planar graphs~\cite{Dross+2014, Dross+2015, Kowalik+2010}, cubic
graphs~\cite{Bondy+1987, Jaeger1974, LiuZhao1996, Punnim2005,
Speckenmeyer1983, Staton1984, ZhengLu1990} and other regular
graphs~\cite{Pike2003, Punnim2006}.

Dreyer and Roberts~\cite{DreyerRoberts2009} have shown that decycling sets in
$r$-regular graphs coincide with~$(r-1)$-conversion sets (see Proposition
\ref{Prop_DR}). Therefore, if~$G$ is~$(k+1)$-regular, then $c_{k}(G)=\phi(G)$.
A detailed survey of results on~$k$-conversion processes, including results on
decycling sets in regular graphs, can be found in~\cite{Wodlinger2018}.

We consider lower bounds on~$c_{k}(G)$ for regular graphs and discuss classes
of graphs that meet, or do not meet, the given bound. We begin, in Section
\ref{section:kk}, by characterizing~$k$-regular graphs having a~$k$-conversion
set of size~$k$. In Section~\ref{kconvnokplus1reg} we consider~$c_{k}(G)$ for
$(k+1)$-regular graphs, first investigating~$(k+1)$-regular graphs with
$c_{k}(G)=k$ and then discussing lower bounds on~$c_{k}(G)$. In Section
\ref{kplusrregular} we obtain a lower bound for~$c_{k}(G)$ for~$(k+r)$-regular
graphs. We restrict our attention to cubic graphs in Section~\ref{Sec_Cubic}
and present classes of cubic graphs that attain the bound for~$c_{2}(G)$, and
others that exceed it. It is known that fullerenes and snarks meet the lower
bound. We study the~$2$-conversion number of graphs that have some of the
defining properties of snarks in Section~\ref{sec:snarks}. Our results in this
section lead us to study~$3$-connected cubic graphs in Section
\ref{sec:3conncubic}, where we construct classes of~$3$-connected cubic graphs~$H_{m}$ of arbitrary girth (and other properties) that exceed the lower bound
for~$c_{2}(H_{m})$ by at least~$m$.

We generally follow the notation of~\cite{ChartrandLesniak}. For graphs~$G$
and~$H$,~$G+H$ denotes the disjoint union of~$G$ and~$H$, and~$G\vee H$
denotes the \emph{join} of~$G$ and~$H$, obtained by adding all possible edges
between~$G$ and~$H$. We denote the independence number of~$G$ by~$\alpha(G)$. 

\section{$k$-Regular graphs with~$k$-conversion number~$k$}

\label{section:kk}

We begin with the straightforward observation that, in order for any
conversion to occur in a~$k$-conversion process, the seed set must contain at
least~$k$ vertices. Therefore~$k$ is a trivial lower bound on~$c_{k}(G)$ for
any graph~$G$ with at least~$k$ vertices. More specifically, if~$G$ is a graph
of order~$n$ and maximum degree~$\Delta$, then~$c_{k}(G)=n$ if~$\Delta<k$ and
otherwise~$c_{k}(G)\geq k$. Leaving aside the case where~$c_{k}(G)=n$, we
focus on graphs with maximum degree at least~$k$ and ask which graphs meet the
bound~$c_{k}(G)=k$.

Graphs that meet this bound are easy to find, and exist for any order~$k+r$,
where~$r\geq1$. (Take, for example, the complete bipartite graph~$K_{k,r}$.)
Imposing structural constraints on~$G$ naturally makes the bound harder to
achieve. In Proposition~\ref{kkk} we give a complete characterization of the
$k$-regular graphs that meet this bound. In Section~\ref{kconvnokplus1reg} we
will expand our investigation of the bound to include~$(k+1)$-regular graphs.

We first state the following proposition by Dreyer and Roberts for referencing.

\begin{propn}
\emph{\cite{DreyerRoberts2009}}\label{Prop_DR}

\begin{enumerate}
\item[(a)] If~$G$ is a~$k$-regular graph, then~$S$ is a~$k$-conversion set
of~$G$ if and only if~$V-S$ is independent.

\item[(b)] If~$G$ is a~$(k+1)$-regular graph, then~$S$ is a~$k$-conversion
set of~$G$ if and only if~$G[V-S]$ is a forest.
\end{enumerate}
\end{propn}

An immediate consequence of Proposition~\ref{Prop_DR}(a) is that if~$G$ is a
$k$-regular graph of order~$n$, then~$c_{k}(G)=n-\alpha(G)$.

\begin{propn}
\label{kkk} A~$k$-regular graph~$G$ has a~$k$-conversion set of size~$k$ (that
is,~$c_{k}(G)=k$) if and only if~$G=H\vee\overline{K_{k-t}}$, where~$H$ is a
$t$-regular graph of order~$k$, and~$0\leq t<k$.
\end{propn}

\noindent\textbf{Proof.\hspace{0.1in}}
Let~$G=H\vee\overline{K_{k-t}}$, where
$H$ and~$t$ are as above. Each vertex of~$\overline{K_{k-t}}$ has~$k$
neighbours in~$H$, so~$V(H)$ is a~$k$-conversion set of size~$k$. Since
vertices of~$\overline{K_{k-t}}$ have no other neighbours, and each vertex of
$H$ has~$t$ neighbours in~$H$ and~$k-t$ neighbours in~$\overline{K_{k-t}}$, $G$ 
is~$k$-regular. For the converse, let~$G$ be a~$k$-regular graph with a
$k$-conversion set~$S$ of order~$k$. By Proposition~\ref{Prop_DR}(a),~$V-S$
is independent. Since~$G$ is~$k$-regular,~$G[S]$ is~$t$-regular for some
$0\leq t<k$ and~$|V-S|=k-t$. The result follows with~$H=G[S]$.
~$\blacksquare$

\section{The~$k$-conversion number of~$(k+1)$-regular graphs}

\label{kconvnokplus1reg}

In this section we present lower bounds on the~$k$-conversion number of a
$(k+1)$-regular graph and determine some properties of the graphs that meet
these bounds. We begin with the trivial lower bound~$c_{k}(G)\geq k$, this time applied to
$(k+1)$-regular graphs. Proposition~\ref{Prop_DR}(b) states that a set~$S$ is a
$k$-conversion set of a~$(k+1)$-regular graph~$G$ if and only if~$G[V-S]$ is
acyclic. In this case~$S$ is also known as a decycling set or a feedback
vertex set. We rely heavily on this characterization of~$k$-conversion sets in
$(k+1)$-regular graphs throughout Section~\ref{kconvnokplus1reg}.

\subsection{$k$-conversion sets of size~$k$ in~$(k+1)$-regular graphs}

\label{kkkplus1}

If~$r\geq1$ and~$G$ is a~$(k+r)$-regular graph with a~$k$-conversion set~$S$
of size~$k$, then every non-seed vertex has at least~$r$ neighbours outside of
$S$. This introduces the possibility that complete conversion of the graph
takes more than one time step. For~$t\geq0$, let~$S_{t}$ be the set of
vertices that convert at time~$t$, starting from a given seed set~$S=S_{0}$.
(It is worth noting that such a graph may still convert in one time step. For
example, consider the~$4$-regular graph~$G=\overline{K_{3}}\vee(K_{2}+K_{2})$,
with~$3$-conversion set~$S=V(\overline{K_{3}})$.)

In Proposition~\ref{nonseedbound} we derive a bound on the number of non-seed vertices in a~$(k+1)$-regular graph with a~$k$-conversion set of size~$k$. We use this result to obtain a sharp upper bound on the order of such a graph (Proposition~\ref{nbound}).

\begin{propn}\label{nonseedbound}
 Let~$G$ be a~$(k+1)$-regular graph and suppose that
$S_{0}$ is a~$k$-conversion set of size~$k$. Then~$|V(G) - S_{0}| <
\frac{k(k+1)-1}{k-1}$.
\end{propn}

\noindent\textbf{Proof.\hspace{0.1in}} We begin by deriving a bound on the number of vertices that convert at time $t=2$ and later. Let~$Y=\cup_{t\geq2}S_{t}$. We count
the edges between~$Y$ and~$S_{0}$ in two ways. First, since~$G$ is
$(k+1)$-regular and each vertex of~$S_{0}$ is adjacent to each vertex of
$S_{1}$, there are at most~$k(k+1-|S_{1}|)$ edges from~$S_{0}$ to~$Y$. On the
other hand, each vertex in~$Y$ has at least~$k$ neighbours that convert before
it. Therefore there are at least~$|Y|k$ edges with at least one endpoint in
$Y$. Since~$G-S_{0}$ is a forest with~$|Y|+|S_{1}|$ vertices, at most
$|Y|+|S_{1}|-1$ have the other endpoint in~$Y\cup S_{1}$. Therefore there are
at least~$|Y|k-|Y|-|S_{1}|+1$ edges from~$Y$ to~$S_{0}$. This gives
$|Y|k-|Y|-|S_{1}|+1\leq k(k+1-|S_{1}|).$ Rearranging, and replacing $Y$ with $\underset{t\ge2}{\cup} S_{t}$, gives the bound 
\begin{equation} \label{Ybound}
\left|  \underset{t\ge2}{\bigcup} S_{t} \right|  \le\frac{ k(k+1) +
|S_{1}|(1-k) -1 }{k-1}.
\end{equation}
The left side of~\eqref{Ybound} equals~$|V(G)-S_{0}|-|S_{1}|$, and the result follows.
~$\blacksquare
$

\bigskip

\begin{figure}[htb]
\begin{center}
\begin{tikzpicture}
\draw (0, 3) -- (0,2); \draw (0,3) -- (2,2); \draw (1, 3) -- (0,2); \draw (1,3) -- (2,2); \draw (2,3) -- (0,2); \draw (2,3) -- (2,2);
\draw (0,1) -- (0,2); \draw (0,1) -- (1,3); \draw (0,1) -- (2,3);
\draw (2,1) -- (2,2); \draw (2,1) -- (0,3); \draw (2,1) -- (1,3);
\draw (1, 0.3) -- (0,1); \draw (1, 0.3) -- (2,1); \draw (1,0.3) -- (0,3); \draw (1, 0.3) -- (2,3);
\draw [fill] (0,3) circle [radius = 0.08];
\draw [fill] (1,3) circle [radius = 0.08];
\draw [fill] (2,3) circle [radius = 0.08];
\draw [fill = white] (0,2) circle [radius =0.08];
\draw [fill = white] (2,2) circle [radius =0.08];
\draw [fill = white] (0,1) circle [radius =0.08];
\draw [fill = white] (2,1) circle [radius =0.08];
\draw [fill = white] (1,0.3) circle [radius =0.08];
\end{tikzpicture}
\end{center}
\caption[A graph that achieves equality in the bound of Proposition~\ref{kbound}.]{A~$4$-regular graph with~$c_3(G) = 3 = |\cup_{t\ge 2}S_t|$, illustrating sharpness of the bound in Proposition~\ref{kbound}. This graph also illustrates the construction in Proposition~\ref{nbound}, with~$k=3$.}
\label{kboundsharp}
\end{figure}
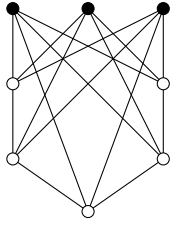

In Proposition~\ref{nbound}, we use Proposition~\ref{nonseedbound} 
to derive an upper bound on the order of a~$(k+1)$-regular graph having 
a~$k$-conversion set of size~$k$ and we prove by construction that the bound is 
sharp for each~$k\geq2$. The result of the construction for~$k=3$ is illustrated in
Figure~\ref{kboundsharp}. Let~$v$ be a vertex such that~$\deg(v)\leq\Delta$.
We define the~$\Delta$-\emph{deficiency} of~$v$ to be~$\operatorname{def}%
_{\Delta}(v)=\Delta-\deg(v)$.

\begin{propn}
\label{nbound} If~$G$ is a~$(k+1)$-regular graph having a~$k$-conversion set
of size~$k$, then the order of~$G$ is at most~$2k+2$. Moreover, for every
$k\geq2$, there exists a~$(k+1)$-regular graph of order~$2k+2$ which has a
$k$-conversion set of size~$k$.
\end{propn}

\noindent\textbf{Proof.\hspace{0.1in}} We obtain the bound for~$k=2$ by
checking all examples (there are three cubic graphs having a~$2$-conversion
set of size~$2$:~$K_{4}$ and the two cubic graphs of order~$6$). For~$k\geq3$,
$\frac{k(k+1)-1}{k-1}>k+3$, so the bound follows from 
Proposition~\ref{nonseedbound}.

To prove that the bound is sharp, we construct a~$(k+1)$-regular graph of
order~$2k+2$ which has a~$k$-conversion set of size~$k$. We begin with the
graph~$K_{2,k}$, where~$S_{0}=S$ is the set of size~$k$ (a~$k$-conversion set)
and~$S_{1}=\{u_{1},v_{1}\}$ is the set of size~$2$ (the set of vertices that
convert at time~$t=1$). For each~$v\in S_{0}$ we now have~$\operatorname{def}%
_{k+1}(v)=k-1$ and for each~$v\in S_{1}$ we have~$\operatorname{def}%
_{k+1}(v)=1$. We will add vertex sets~$S_{2},S_{3},\dots$ such that the
vertices of~$S_{i}$ convert at time~$t=i$ from the~$k$-conversion set~$S_{0}$.
To achieve this, for each~$i\geq2$, we must add at least~$k$ edges from
$S_{i}$ to~$\cup_{j=0}^{i-1}S_{j}$. Some care is required in choosing the
edges, in order to ensure that there will always be at least~$k$ distinct
vertices available in~$\cup_{j=0}^{i-1}S_{j}$. For~$i\geq2$, if there are
still at least~$k-1$ vertices in~$S_{0}$ of deficiency at least 2, let
$S_{i}=\{u_{i},v_{i}\}$. Join~$u_{i}$ to~$u_{i-1}$ and to~$k-1$ vertices of
$S_{0}$, beginning with those of highest deficiency. Then join~$v_{i}$ to
$v_{i-1}$ and to~$k-1$ vertices of~$S_{0}$, once again beginning with those of
highest deficiency. Joining~$u_{i}$ and~$v_{i}$ to~$u_{i-1}$ and~$v_{i-1}$ at
each step means that the vertices of~$S_{1},\dots,S_{i-1}$ have degree~$k+1$,
so the only deficient vertices are the newly added ones and those in~$S_{0}$.
Joining the new vertices first to the vertices of highest deficiency in
$S_{0}$ guarantees that the deficiencies among the vertices of~$S_{0}$ are
always within 1 of each other. Therefore, the first time there fail to be at
least~$k-1$ vertices in~$S_{0}$ with deficiency at least~$2$, there are either
no deficient vertices in~$S_{0}$ (if~$k$ is even) or there are~$k-1$ deficient
vertices in~$S_{0}$ and their deficiency is~$1$ (if~$k$ is odd). In the case
where~$k$ is even, we add vertices~$u_{i}$ and~$v_{i}$~$\frac{k}{2}$ times
before we run out of deficient vertices in~$S_{0}$. That is, the process stops
when~$i=\frac{k}{2}+1$, and~$|\cup_{i=2}^{\frac{k}{2}+1}S_{i}|=k$. Adding an
edge between~$u_{\frac{k}{2}+1}$ and~$v_{\frac{k}{2}+1}$ yields a simple
$(k+1)$-regular graph of order~$2k+2$ (including the~$k$ vertices of~$S_{0}$
and the 2 vertices of~$S_{1}$). In the case where~$k$ is odd, we add
$\frac{k-1}{2}$ pairs of vertices~$u_{i}$ and~$v_{i}$ before the deficiencies
in~$S_{0}$ become too small. That is, the process stops when~$i=\frac{k+1}{2}$
and~$|\cup_{i=2}^{\frac{k+1}{2}}S_{i}|=k-1$. We complete the~$(k+1)$-regular
graph by adding one final vertex,~$w$, and joining it to~$u_{\frac{k+1}{2}}$,
$v_{\frac{k+1}{2}}$ and to the~$k-1$ vertices of deficiency~$1$ in~$S_{0}$.
The total number of vertices is now~$2k+2$, including the~$k$ vertices of~$S_{0}$ and 
the~$2$ vertices of~$S_{1}$.~$\blacksquare$

\bigskip

In the proof of Proposition~\ref{nonseedbound}, we derived the bound~\eqref{Ybound} on the size of $\underset{t\geq2}{\cup}S_{t}$ for~$(k+1)$-regular graphs with a
$k$-conversion set of size~$k$. Proposition~\ref{kbound}, below, provides another upper bound on the same quantity. 
When~$|S_{1}|\geq\frac{2k-1}{k-1}$, the bound
provided by~\eqref{Ybound} 
 is stronger than that of Proposition~\ref{kbound}. 
However, the bound of Proposition~\ref{kbound} 
is sharp for
small values of~$|S_{1}|$, as shown by the graph in Figure~\ref{kboundsharp}.

\begin{propn}
\label{kbound} Let~$G$ be a~$(k+1)$-regular graph with a~$k$-conversion set of
size~$k$. Then $\big|\underset{t\ge2}{\cup} S_{t} \big| \le k$.
\end{propn}

\noindent\textbf{Proof.\hspace{0.1in}} Let~$Y=\cup_{t\geq2}S_{t}$. By
Proposition~\ref{Prop_DR}(b),~$G-S_{0}$ is a forest~$F$, and its leaves are the
vertices in~$S_{1}$. Therefore, for every~$v\in Y$,~$\deg_{F}(v)\leq|S_{1}|$,
and~$\deg_{G}(v)=k+1$, so~$v$ has at least~$k+1-|S_{1}|$ neighbours in~$S_{0}%
$. Hence there are at least~$|Y|(k+1-|S_{1}|)$ edges between~$Y$ and~$S_{0}$.
On the other hand, there are at most~$k(k+1-|S_{1}|)$ edges between~$S_{0}$
and~$Y$, by the argument given in the proof of Proposition~\ref{nonseedbound}. 
Therefore~$|Y|(k+1-|S_{1}|)\leq k(k+1-|S_{1}|)$.~$\blacksquare$

\subsection{A lower bound on~$c_{k}(G)$ for~$(k+1)$-regular graphs}

\label{sec:regularlowerbound}

In Sections~\ref{section:kk} and~\ref{kkkplus1} we began with a fixed seed set
size (namely~$k$, the minimum possible size for a nontrivial~$k$-conversion
set), and asked which graphs have a~$k$-conversion set of this size. We
obtained constraints on the structure and order (respectively) of~$k$- and
$(k+1)$-regular graphs with this property. In this section we instead begin
with a class of graphs, and ask how small a~$k$-conversion set can be for a
graph in this class.

As discussed in Section~\ref{section:kk},~$k$ is a lower bound on the
$k$-conversion number of any graph with order at least~$k$. While it is
possible to have arbitrarily large graphs that attain this bound, for many
classes of graphs a~$k$-conversion set of size~$k$ can only convert a limited
number of vertices. Indeed, we showed in Proposition~\ref{nbound} that in the
class of~$(k+1)$-regular graphs, a~$k$-conversion set of size~$k$ can convert
at most~$2k+2$ vertices. For these graphs, as the order grows beyond the
$2k+2$ threshold, we require more than~$k$ seed vertices to convert the graph.
In this case,~$k$ is no longer a good lower bound for the~$k$-conversion number.

Beinecke and Vandell~\cite[Corollary 1.2]{BeinekeVandell1997} showed that if
$G$ is a graph with~$n$ vertices,~$m$ edges and maximum degree~$\Delta$, then
the decycling number of~$G$ is at least~$\frac{m-n+1}{\Delta-1}$. This
generalized the lower bound obtained by Staton~\cite{Staton1984} on the
decycling number of~$(k+1)$-regular graphs, which corresponds to the
$k$-conversion number. We present a proof of Staton's result which yields a
condition for equality in the bound.

\begin{propn}
\label{regularlowerbound} Let~$G$ be a~$(k+1)$-regular graph of order~$n$,
$k\ge2$. Then~$c_{k}(G) \ge\left\lceil \frac{n(k-1)+2}{2k}\right\rceil$.
Moreover, a minimum~$k$-conversion set~$S$ of~$G$ has size~$\frac
{n(k-1)+2}{2k}$ if and only if~$S$ is independent and~$G-S$ is a tree.
\end{propn}

\noindent\textbf{Proof.\hspace{0.1in}} Let~$S$ be a minimum~$k$-conversion set
of~$G$, and let~$\overline{S}=V(G) - S$. For~$X\in\{S,\overline{S}\}$, let
$n_{X}$ and~$m_{X}$ denote the number of vertices and edges, respectively, in
$G[X]$. Counting in two ways the number of edges between~$S$ and~$\overline
{S}$ gives the identity
\[
(k+1)n_{S}-2m_{S}=(k+1)n_{\overline{S}}-2m_{\overline{S}}.
\]
By Proposition~\ref{Prop_DR}(b),~$G[\overline{S}]$ is a forest; let~$y$ be
its number of components. Then
\[
(k+1)n_{S}-2m_{S}=(k+1)n_{\overline{S}}-2(n_{\overline{S}}-y).
\]
Substituting~$n_{\overline{S}}=n-n_{S}$, and rearranging, this gives
\[
n_{S}=\frac{n(k-1)+2m_{s}+2y}{2k}.
\]
Therefore,~$c_{k}(G)=n_{S}\geq\frac{n(k-1)+2}{2k}$, with equality if and only
if~$S$ is independent and~$G-S$ is a tree. In particular,~$c_{k}%
(G)\geq\left\lceil \frac{n(k-1)+2}{2k}\right\rceil$.~$\blacksquare$

\bigskip

In the next section we prove a lower bound similar to that of Proposition
\ref{regularlowerbound} for~$(k+r)$-regular graphs.

\section{A lower bound on~$c_{k}(G)$ for~$(k+r)$-regular graphs}

\label{kplusrregular}

Dreyer and Roberts~\cite{DreyerRoberts2009} give a lower bound of
$\frac{(k-r)n}{2k}$ on~$c_{k}(G)$ for~$(k+r)$-regular graphs of order~$n$, for
$0\leq r<k$. In the case~$r=k-1$, where~$G$ is a~$(2k-1)$-regular graph,
Zaker~\cite{Zaker2012} strengthens this bound to~$c_{k}(G)\geq\frac
{n+2(k-1)}{2k}$.
In this section we improve upon both of these previous bounds by providing, in
Proposition~\ref{kplusrreglowerbound}, a new lower bound of~$c_{k}(G)\geq
\frac{(k-r)n+(r+1)r}{2k}$, which is sharp for all $r$,~$0\leq r\le k-1$.

Proposition~\ref{kplusrimmunesets} generalizes Proposition~\ref{Prop_DR} by
characterizing the~$k$-conversion sets~$S$ of~$(k+r)$-regular graphs in terms
of a condition on~$V-S$. For~$r\geq0$, a graph~$G$ is~$r$-\emph{degenerate} if
every induced subgraph of~$G$ has a vertex of degree at most~$r$. We say
that~$G$ is a \emph{maximal }$r$\emph{-degenerate} graph if~$G$ is
$r$-degenerate but for every pair of non-adjacent vertices~$x,y$ in~$G$,
adding the edge~$xy$ to~$E(G)$ produces a graph that is not~$r$-degenerate. We
note that a graph~$G$ is~$0$-degenerate if and only if it has no edges, and it
is~$1$-degenerate if and only if it is acyclic.

We call a nonempty set $U$ of vertices of a graph $G$ \emph{$k$-immune} if every vertex in $U$ has fewer than $k$ neighbours in $V - U$. It is straightforward to see that $S \subseteq V$ is a $k$-conversion set of $G$ if and only if $V -S$ does not contain a $k$-immune set. We use this observation in the proof of Proposition \ref{kplusrimmunesets}, and again in Section~\ref{sec:snarks}.

\begin{propn}
\label{kplusrimmunesets} Let~$G$ be a~$(k+r)$-regular graph, with~$r \ge0$. A
set~$S$ of vertices of~$G$ is a~$k$-conversion set if and only if~$G[V-S]$ is
$r$-degenerate.
\end{propn}

\noindent\textbf{Proof.\hspace{0.1in}} Suppose~$V-S$ is~$r$-degenerate, so
every subgraph~$H$ of~$V-S$ has a vertex of degree at most~$r$. In other
words, some vertex of~$H$ has at least~$k$ neighbours in~$G-H$. Let
$H_{0}=V-S$ and let~$S_{1}$ be the set of vertices of degree at most~$r$ in
$H_{0}$. These vertices have at least~$k$ neighbours in~$G-H_{0}=S$, so they
convert at time~$t=1$. Let~$H_{1}=H_{0}-S_{1}$ and let~$S_{2}$ be the set of
vertices of degree at most~$r$ in~$H_{1}$. These vertices have at least~$k$
neighbours in~$V-H_{1}=S\cup S_{1}$, so they convert at time~$t=2$. Continue
this process until some~$H_{i}=\emptyset$. At each step, the set~$V-H_{j}$ is
converted, so when~$H_{i}=\emptyset$ the whole graph is converted. Therefore
$S$ is a~$k$-conversion set. On the other hand, if~$V-S$ is not~$r$-degenerate
then there is some subgraph~$H$ of~$V-S$ in which no vertex has~$k$ neighbours
outside~$H$. 
Therefore~$V(H)$ is a~$k$-immune set, so~$S$ is not a conversion set
of~$G$.~$\blacksquare$

\bigskip

Proposition~\ref{kplusrreglowerbound} generalizes Proposition
\ref{regularlowerbound}, establishing a lower bound on~$c_{k}(G)$ for~$\left
(k+r \right)$-regular graphs~$G$. The proof technique is the same as for
Proposition~\ref{regularlowerbound}, but requires the following lemma, due to
Lick and White, bounding the number of edges in an~$r$-degenerate graph.

\begin{lemma}
\emph{\cite[Proposition 3 and Corollary 1]{LickWhite1970}}
\label{degenerateedgecount} Let~$G$ be an~$r$-degenerate graph with~$n \ge r$
vertices and~$m$ edges. Then~$m \le rn - \binom{r+1}{2}$, with equality if and
only if~$G$ is maximal~$r$-degenerate.
\end{lemma}

\begin{propn}
\label{kplusrreglowerbound} Let~$G$ be a~$(k+r)$-regular graph of order~$n$,
where~$0\le r < k$. Then
\[
c_{k}(G) \ge\frac{(k-r)n + (r+1)r}{2k}.
\]
Moreover, for~$r\ge1$, a minimum~$k$-conversion set~$S$ of~$G$ has order
$\frac{(k-r)n + (r+1)r}{2k}$ if and only if~$S$ is independent and~$G[V-S]$ is
a maximal~$r$-degenerate graph.

\end{propn}

\noindent\textbf{Proof.\hspace{0.1in}} First suppose~$r=0$. Proposition
\ref{Prop_DR}(a) implies that~$c_{k}(G)=n-\alpha(G)$. Since~$G$ is regular,
$\alpha(G)\leq\frac{n}{2}$, and the result follows. Now let~$r\geq1$ and let
$G$ be a~$(k+r)$-regular graph with~$n>k+r$ vertices. Let~$S$ be a
$k$-conversion set of~$G$ and for~$X\in\{S,\overline{S}\}$, let~$n_{X}$ and
$m_{X}$ denote the number of vertices in~$X$ and the number of edges in
$G[X]$, respectively. Counting in two ways the edges between~$S$ and
$\overline{S}$ gives
\[
(k+r)n_{S}-2m_{S}=(k+r)n_{\overline{S}}-2m_{\overline{S}}.
\]
Applying the bound~$m_{\overline{S}}\leq rn_{\overline{S}}-\binom{r+1}{2}$, as
provided by Lemma~\ref{degenerateedgecount}, and simplifying gives
\[
(k+r)n_{S}-2m_{S}\geq(k-r)n_{\overline{S}}+(r+1)r,
\]
with equality if and only if~$G[\overline{S}]$ is maximal~$r$-degenerate. By
substituting~$n_{\overline{S}}=n-n_{S}$ and rearranging, we obtain
\[
n_{S}\geq\frac{(k-r)n+(r+1)r+2m_{S}}{2k},
\]
with equality if and only if~$G[\overline{S}]$ is maximal~$r$-degenerate. The
result follows since~$m_{S}\geq0$ with equality if and only if~$S$ is an
independent set.~$\blacksquare$

\bigskip

We note that, by definition of maximal~$r$-degeneracy, in order to determine
whether a subgraph~$H$ of~$G$ (in particular,~$H=G[\overline{S}]$) is maximal
$r$-degenerate we must look at all~$x,y\in V(H)$ such that~$xy\not \in
E(H)$---regardless of whether~$xy\in E(G)$---and determine whether~$H+xy$ is
still~$r$-degenerate. In other words the maximality of~$H$ with respect to
$r$-degeneracy does not depend on whether we can add more vertices or edges of
$G$ into~$H$ without losing the~$r$-degenerate property, but whether we can
add an edge between two non-adjacent vertices of~$H$. In particular, when
$H=G[\overline{S}]$,~$H$ is an induced subgraph so any additional edge~$xy$
under consideration is necessarily absent from~$G$.

\section{Cubic graphs}

\label{Sec_Cubic}For~$k=2$, Proposition~\ref{regularlowerbound} gives the
lower bound
\begin{equation}
c_{2}(G)\geq\left\lceil \frac{n+2}{4}\right\rceil \label{cubiclowerboundceil}%
\end{equation}
for cubic graphs~$G$ of order~$n$. In this section we present classes of cubic
graphs that attain this bound and others that exceed it. We begin by stating a
result by Payan and Sakarovitch~\cite{PayanSakarovitch} that provides a
sufficient condition for equality in the bound. 

A graph~$G$ is \emph{cyclically }$k$\emph{-edge connected} (\emph{cyclically
}$k$\emph{-vertex connected}) if at least~$k$ edges (vertices) must be removed
to disconnect~$G$ into two subgraphs that each contain a cycle. A cubic graph
$G\notin\{K_{4},K_{3,3}\}$ is cyclically~$4$-edge connected if and only if it
is cyclically~$4$-vertex connected~\cite{McCuaig1992}, so we simply call such
graphs \emph{cyclically }$4$\emph{-connected}.

\begin{theorem}
\emph{\cite{PayanSakarovitch}}\label{Thm_4-conn}\ If~$G$ is a cyclically
$4$-connected cubic graph of order~$n$, then%
\[
c_{2}(G)=\left\lceil \frac{n+2}{4}\right\rceil.
\]

\end{theorem}

A \emph{fullerene} is a planar cubic graph whose faces, including the outer
face, in any plane representation, all have size~$5$ or~$6$. Do{\v{s}}%
li{\'{c}}~\cite[Theorem 8]{Doslic1998} proved that all fullerenes are
cyclically~$4$-edge connected, and therefore by Theorem~\ref{Thm_4-conn} they
achieve equality in the lower bound~\eqref{cubiclowerboundceil}. 

\subsection{Snarks and would-be snarks}
\label{sec:snarks}

By Vizing's theorem~\cite[Theorem 17.2]{ChartrandLesniak},
if~$G$ is a graph with maximum degree~$\Delta$, then~$G$ has chromatic index
$\Delta$ or~$\Delta+1$; in the former case,~$G$ is of Class~$1$, and in the
latter case, of Class~$2$.

A \emph{snark} is a connected, bridgeless, Class~$2$ cubic graph. To avoid
degenerate cases, it has long been standard to require snarks to be
triangle-free. They have been studied since the 1880's, when Tait
\cite{Tait1880} proved that the Four Colour Theorem is equivalent to the
statement that no snark is planar. We refer to such snarks (connected,
bridgeless, triangle-free Class~$2$ cubic graphs) as \emph{Gardner snarks}, as
this was the common definition of snarks when Martin Gardner gave them the
name \textquotedblleft snark\textquotedblright\ in 1975~\cite{Gardner1976}.
The name, taken from the elusive creature in Lewis Carroll's poem \emph{The
Hunting of the Snark}, reflects the scarcity of examples in the years after
Tait defined them. The smallest and earliest known example of a snark is the
Petersen graph, first mentioned by Alfred Bray Kempe\ in 1886~\cite{Kempe} and
named after the Danish mathematician Julius Petersen,\ who presented it as
counterexample to Tait's claim that all cubic graphs were 3-edge colourable.
Due to their connection with the Four Colour Theorem (Four Colour Conjecture,
at the time), much attention was given to the pursuit of new examples of
snarks (with the hope of finding a planar one, perhaps), but a second example
was not discovered until 1946. Since then, more examples have been discovered,
including infinite families.

Interest in snarks has remained steady, due in part to their connection to
other important conjectures in graph theory, notably the Cycle Double Cover
Conjecture~\cite{DeVos2007, Szekeres1973}. In 1985, Jaeger~\cite{Jaeger1985}
proved that a smallest counterexample to the conjecture must be a snark;
therefore, if the conjecture is true for snarks, it is true for all graphs.

More recently, more restrictive definitions of snarks have become the
standard. It is now common to require snarks to have higher connectivity and
larger girth. Some authors use even more restrictive definitions in order to
exclude snarks that can be obtained from other snarks. Some require them to be
cyclically~$4$-edge connected, rather than simply triangle-free~\cite{HouseOfGraphs}. We call cyclically~$4$-edge connected snarks of girth at
least five (at least four) \emph{strong (weak) snarks}. A convenient overview
of approximately the first century of snark research, including a discussion
of modern definitions, can be found in~\cite{Watkins1989}. 

By Theorem~\ref{Thm_4-conn}, strong and weak snarks achieve equality in the
lower bound~\eqref{cubiclowerboundceil}. It is therefore natural to ask
whether all snarks do. However, we will show in Section~\ref{sec:3conncubic}
that there exist infinitely many Gardner snarks that fail to meet the bound. 

Theorems~\ref{planarimpliesclass1} and~\ref{bridgedimpliesclass2} give
well-known sufficient conditions for cubic graphs to be Class~$1$ and Class~$2$, 
 respectively, which aids our search for examples in each category.
Theorem~\ref{planarimpliesclass1} was shown by Tait in 1880 to be equivalent
to the Four Colour Theorem.

\begin{theorem}
\label{planarimpliesclass1}\emph{\cite{AppelHaken1977, Appel+1977, Tait1880}}
Every bridgeless planar cubic graph has chromatic index~$3$.
\end{theorem}

\begin{theorem}
\label{bridgedimpliesclass2}\emph{\cite{Isaacs}}\  Every bridged cubic graph
has chromatic index~$4$.
\end{theorem}

Theorem~\ref{bridgedimpliesclass2} allows us to limit our investigation to
graphs that are bridgeless or Class~$2$, since there are no bridged, Class~$1$
cubic graphs. All other combinations---that is, all allowable
combinations---of the three defining characteristics of snarks (bridgeless,
Class~$2$, triangle-free) admit graphs that meet the lower bound and graphs
that do not meet the lower bound. Table~\ref{table:snarkpropertycombos} gives
an example of a graph for each type for each of the combinations.

\newcolumntype{L}[1]{>{\raggedright\let\newline\\\arraybackslash\hspace{0pt}}m{#1}}
\newcolumntype{C}[1]{>{\centering\let\newline\\\arraybackslash\hspace{0pt}}m{#1}}
\newcolumntype{K}[1]{>{\centering\let\newline\\\arraybackslash\hspace{0pt}}m{#1}}
\newcolumntype{R}[1]{>{\raggedleft\let\newline\\\arraybackslash\hspace{0pt}}m{#1}}

\begin{table}
  \centering
  \begin{threeparttable}[c]
  \begin{tabular}{@{} C{1.4cm}  C{1.4cm}  C{1.4cm}  C{4.4cm}  C{4.4cm} @{}}
    \toprule
  Bridgeless?& Class 2? &~$\Delta$-free? & Example with~$c_2(G) = \left \lceil \frac{n+2}{4} \right \rceil$ & Example with~$c_2(G) > \left \lceil \frac{n+2}{4} \right \rceil$ \\ [1.4ex]
    \hline
\\ [-0.2ex]
    No & Yes & No  & \vspace{-6mm}\begin{center}
\begin{tikzpicture}[scale = 0.3,  
white_vertex/.style={whitevertex,inner sep=0pt,minimum size=0.12cm},
black_vertex/.style={blackvertex,inner sep=0pt,minimum size=0.12cm},
solid_edge/.style = {thick, black},
wavy_edge/.style={thick, black, decorate, decoration={snake, segment length = 3mm, amplitude = 0.6mm}},
dotted_edge/.style={thick, black, dashed}]


\begin{scope}
\foreach \x/\y in {0/1, 72/2, 144/3,  216/4, 288/5}{ 
  	\node[white_vertex] (\y) at (canvas polar cs: radius = 1.7cm, angle = \x){};
}
\end{scope}

\draw[solid_edge](1)--(2)--(3) --(4) --(5)--(1);
\draw[solid_edge](2)--(4);
\draw[solid_edge](3)--(5);

\begin{scope}[xshift = 4.6cm, rotate = 180]
\foreach \y/\x in {6/0, 7/51.4, 8/102.8, 9/154.3, 10/205.7, 11/257.1, 12/308.6}{
	\node[white_vertex] (\y) at (canvas polar cs: radius = 2cm, angle = \x){};
}
\end{scope}

\draw[solid_edge] (6)--(7)--(8)--(9)--(10)--(11)--(12)--(6);
\draw[solid_edge] (7)--(10);
\draw[solid_edge] (8)--(11);
\draw[solid_edge] (9) -- (12);

\draw[solid_edge] (1)--(6);

\end{tikzpicture}
\end{center} & \vspace{-6mm}\begin{center}
\begin{tikzpicture}[scale = 0.3,  
white_vertex/.style={whitevertex,inner sep=0pt,minimum size=0.12cm},
black_vertex/.style={blackvertex,inner sep=0pt,minimum size=0.12cm},
solid_edge/.style = {thick, black},
wavy_edge/.style={thick, black, decorate, decoration={snake, segment length = 3mm, amplitude = 0.6mm}},
dotted_edge/.style={thick, black, dashed}]


\begin{scope}
\foreach \x/\y in {0/1, 72/2, 144/3,  216/4, 288/5}{ 
  	\node[white_vertex] (\y) at (canvas polar cs: radius = 1.7cm, angle = \x){};
}
\end{scope}

\draw[solid_edge](1)--(2)--(3) --(4) --(5)--(1);
\draw[solid_edge](2)--(4);
\draw[solid_edge](3)--(5);

\begin{scope}[xshift = 4.6cm, rotate = 180]
\foreach \x/\y in {0/6, 72/7, 144/8,  216/9, 288/10}{
	\node[white_vertex] (\y) at (canvas polar cs: radius = 1.7cm, angle = \x){};
}
\end{scope}

\draw[solid_edge] (6)--(7)--(8)--(9)--(10)--(6);
\draw[solid_edge] (7)--(9);
\draw[solid_edge] (8)--(10);

\draw[solid_edge] (1)--(6);

\end{tikzpicture}
\end{center} \\  [0.4ex]
    
    No & Yes & Yes & \vspace{-6mm}\begin{center}
\begin{tikzpicture}[scale = 0.3,  
white_vertex/.style={whitevertex,inner sep=0pt,minimum size=0.12cm},
black_vertex/.style={blackvertex,inner sep=0pt,minimum size=0.12cm},
solid_edge/.style = {thick, black},
wavy_edge/.style={thick, black, decorate, decoration={snake, segment length = 3mm, amplitude = 0.6mm}},
dotted_edge/.style={thick, black, dashed}]


\begin{scope}
\foreach \y/\x in {1/0, 2/51.4, 3/102.8, 4/154.3, 5/205.7, 6/257.1, 7/308.6}{  
  	\node[white_vertex] (\y) at (canvas polar cs: radius = 2cm, angle = \x){};
}
\end{scope}

\draw[solid_edge](1)--(2)--(3) --(4) --(5)--(6)--(7)--(1);
\draw[solid_edge](2)--(5);
\draw[solid_edge](3)--(6);
\draw[solid_edge](4)--(7);

\begin{scope}[xshift = 5cm, rotate = 180]
\foreach \y/\x in {8/0, 9/51.4, 10/102.8, 11/154.3, 12/205.7, 13/257.1, 14/308.6}{
	\node[white_vertex] (\y) at (canvas polar cs: radius = 2cm, angle = \x){};
}
\end{scope}

\draw[solid_edge] (8)--(9)--(10)--(11)--(12)--(13)--(14)--(8);
\draw[solid_edge] (9)--(12);
\draw[solid_edge] (10)--(13);
\draw[solid_edge] (11) -- (14);

\draw[solid_edge] (1)--(8);

\end{tikzpicture}
\end{center} & Any triangle-free cubic graph of the form \vspace{3mm}\begin{tikzpicture}[baseline=-0.5ex]
    \path (0,0) node[circle, inner sep=1, draw](x) {$H$}
    		(1,0) node[circle, inner sep=1, draw](y) {$H$};
	\draw (x)--(y);
    \end{tikzpicture} \vspace{-3mm}where H has order~$n \equiv 1 \pmod 4$ \\ [7ex] 
    
    Yes & No & No & \vspace{-7mm}\begin{center}
\begin{tikzpicture}[scale = 0.3,  
white_vertex/.style={whitevertex,inner sep=0pt,minimum size=0.12cm},
black_vertex/.style={blackvertex,inner sep=0pt,minimum size=0.12cm},
solid_edge/.style = {thick, black},
wavy_edge/.style={thick, black, decorate, decoration={snake, segment length = 3mm, amplitude = 0.6mm}},
dotted_edge/.style={thick, black, dashed}]


\begin{scope}[rotate=-30]
\foreach \x/\y in {0/1, 120/2, 240/3}{ 
  	\node[white_vertex] (\y) at (canvas polar cs: radius = 1cm, angle = \x){};
}
\foreach \x/\y in {0/4, 120/5, 240/6}{ 
  	\node[white_vertex] (\y) at (canvas polar cs: radius = 2.5cm, angle = \x){};
}
\end{scope}

\draw[solid_edge](1)--(2)--(3)--(1);
\draw[solid_edge](4)--(5)--(6)--(4);
\draw[solid_edge](2)--(5);
\draw[solid_edge] (3)--(6);

\begin{scope}[xshift = 6cm, rotate = 210]
\foreach \x/\y in {0/7, 120/8, 240/9}{ 
  	\node[white_vertex] (\y) at (canvas polar cs: radius = 1cm, angle = \x){};
}
\foreach \x/\y in {0/10, 120/11, 240/12}{ 
  	\node[white_vertex] (\y) at (canvas polar cs: radius = 2.5cm, angle = \x){};
}
\end{scope}

\draw[solid_edge](7)--(8)--(9)--(7);
\draw[solid_edge](10)--(11)--(12)--(10);
\draw[solid_edge] (8)--(11);
\draw[solid_edge] (9)--(12);

\draw[solid_edge] (1)--(7);
\draw[solid_edge] (4) -- (10);

\end{tikzpicture}
\end{center} & 
    \vspace{-7mm}\begin{center}
\begin{tikzpicture}[scale = 0.35,  
point/.style={blackvertex,inner sep=0pt,minimum size=0cm},
white_vertex/.style={whitevertex,inner sep=0pt,minimum size=0.12cm},
black_vertex/.style={blackvertex,inner sep=0pt,minimum size=0.12cm},
solid_edge/.style = {thick, black},
wavy_edge/.style={thick, black, decorate, decoration={snake, segment length = 3mm, amplitude = 0.6mm}},
dotted_edge/.style={thick, black, dashed}]

\begin{scope}[xscale = 0.8, yscale = 0.6] 
\foreach \x/\y in {0/11, 60/12, 120/13,  180/14, 240/15, 300/16}{ 
  	\node[white_vertex] (\y) at (canvas polar cs: radius = 1.7cm, angle = \x){};
}
\draw[solid_edge](11)--(12)--(13) --(14) --(15)--(16)--(11);
\draw[solid_edge](12)--(16);
\draw[solid_edge](13)--(15);
\end{scope}

\begin{scope}[ xshift = 3.5cm, yshift = -2cm, rotate=-60,xscale = 0.8, yscale=0.6,] 
\foreach \x/\y in {0/21, 60/22, 120/23,  180/24, 240/25, 300/26}{ 
  	\node[white_vertex] (\y) at (canvas polar cs: radius = 1.7cm, angle = \x){};
}
\draw[solid_edge](21)--(22)--(23) --(24) --(25)--(26)--(21);
\draw[solid_edge](22)--(26);
\draw[solid_edge](23)--(25);
\end{scope}

\begin{scope}[xshift = -3.5cm, yshift = -2cm, rotate=60, xscale = 0.8,yscale=0.6, ] 
\foreach \x/\y in {0/31, 60/32, 120/33,  180/34, 240/35, 300/36}{ 
  	\node[white_vertex] (\y) at (canvas polar cs: radius = 1.7cm, angle = \x){};
}
\draw[solid_edge](31)--(32)--(33) --(34) --(35)--(36)--(31);
\draw[solid_edge](32)--(36);
\draw[solid_edge](33)--(35);
\end{scope}


\node[point] (a) at (-3cm, -4.5cm){};
\node[point] (b) at (3cm, -4.5cm){};

\draw[dotted_edge] (a) to [out = -10, in = 190] (b);
\draw[solid_edge] (34) to [out = -80, in = 170] (a);
\draw[solid_edge] (b) to [out = 10, in = 260] (21);

\draw[solid_edge](31)--(14);
\draw[solid_edge](11)--(24);

\end{tikzpicture}
\end{center} \\ [0.2ex] 
    
    Yes & No & Yes &~$Q_3$, Fullerenes &\vspace{-5mm} \begin{center}
\begin{tikzpicture}[scale = 0.35,  
point/.style={blackvertex,inner sep=0pt,minimum size=0cm},
white_vertex/.style={whitevertex,inner sep=0pt,minimum size=0.12cm},
black_vertex/.style={blackvertex,inner sep=0pt,minimum size=0.12cm},
solid_edge/.style = {thick, black},
wavy_edge/.style={thick, black, decorate, decoration={snake, segment length = 3mm, amplitude = 0.6mm}},
dotted_edge/.style={thick, black, dashed}]

\begin{scope}[xscale = 0.8, yscale = 0.6] 
\foreach \x/\y in {0/11, 60/12, 120/13,  180/14, 240/15, 300/16}{ 
  	\node[white_vertex] (\y) at (canvas polar cs: radius = 1.7cm, angle = \x){};
}
\draw[solid_edge](11)--(12)--(13) --(14) --(15)--(16)--(11);
\draw[solid_edge](12)--(15);
\draw[solid_edge](13)--(16);
\end{scope}

\begin{scope}[ xshift = 3.5cm, yshift = -2cm, rotate=-60,xscale = 0.8, yscale=0.6,] 
\foreach \x/\y in {0/21, 60/22, 120/23,  180/24, 240/25, 300/26}{ 
  	\node[white_vertex] (\y) at (canvas polar cs: radius = 1.7cm, angle = \x){};
}
\draw[solid_edge](21)--(22)--(23) --(24) --(25)--(26)--(21);
\draw[solid_edge](22)--(25);
\draw[solid_edge](23)--(26);
\end{scope}

\begin{scope}[xshift = -3.5cm, yshift = -2cm, rotate=60, xscale = 0.8,yscale=0.6, ] 
\foreach \x/\y in {0/31, 60/32, 120/33,  180/34, 240/35, 300/36}{ 
  	\node[white_vertex] (\y) at (canvas polar cs: radius = 1.7cm, angle = \x){};
}
\draw[solid_edge](31)--(32)--(33) --(34) --(35)--(36)--(31);
\draw[solid_edge](32)--(35);
\draw[solid_edge](33)--(36);
\end{scope}


\node[point] (a) at (-3cm, -4.5cm){};
\node[point] (b) at (3cm, -4.5cm){};

\draw[dotted_edge] (a) to [out = -10, in = 190] (b);
\draw[solid_edge] (34) to [out = -80, in = 170] (a);
\draw[solid_edge] (b) to [out = 10, in = 260] (21);

\draw[solid_edge](31)--(14);
\draw[solid_edge](11)--(24);

\end{tikzpicture}
\end{center}\\ [-1.5ex] 
    
    Yes & Yes & No & \vspace{-3mm}\begin{center}
\begin{tikzpicture}[scale = 0.3,  
white_vertex/.style={whitevertex,inner sep=0pt,minimum size=0.12cm},
black_vertex/.style={blackvertex,inner sep=0pt,minimum size=0.12cm},
solid_edge/.style = {thick, black},
wavy_edge/.style={thick, black, decorate, decoration={snake, segment length = 3mm, amplitude = 0.6mm}},
dotted_edge/.style={thick, black, dashed}]


\begin{scope}[rotate = 90]
\foreach \x/\y in {0/1, 72/2, 144/3,  216/4, 288/5}{ 
  	\node[white_vertex] (\y) at (canvas polar cs: radius = 1.2cm, angle = \x){};
}
\foreach \x/\y in {0/6, 40/7, 80/8,  120/9, 160/10, 200/11, 240/12, 280/13, 320/14}{ 
  	\node[white_vertex] (\y) at (canvas polar cs: radius = 2.5cm, angle = \x){};
}
	
\draw[solid_edge](1)--(3)--(5) --(2) --(4)--(1);
\draw[solid_edge](6)--(7)--(8)--(9)--(10)--(11)--(12)--(13)--(14)--(6);
\draw[solid_edge](1)--(6);
\draw[solid_edge](2)--(8);
\draw[solid_edge](3)--(10);
\draw[solid_edge](4)--(11);
\draw[solid_edge](5)--(13);
\draw[solid_edge](7)--(9);
\draw[solid_edge](12)--(14);

\end{scope}

\end{tikzpicture}
\end{center} & \begin{center}
\begin{tikzpicture}[scale = 0.3,  
white_vertex/.style={whitevertex,inner sep=0pt,minimum size=0.12cm},
black_vertex/.style={blackvertex,inner sep=0pt,minimum size=0.12cm},
solid_edge/.style = {thick, black},
wavy_edge/.style={thick, black, decorate, decoration={snake, segment length = 3mm, amplitude = 0.6mm}},
dotted_edge/.style={thick, black, dashed}]


\begin{scope}[rotate = 36]
\foreach \x/\y in {0/1, 30/2, 72/3, 102/4, 144/5, 174/6,  216/7, 246/8, 288/9, 318/10}{ 
  	\node[white_vertex] (\y) at (canvas polar cs: radius = 3.8cm, angle = \x){};
}
\foreach \x/\y in {15/11, 87/12, 159/13, 231/14, 303/15}{   
	\node[white_vertex] (\y) at (canvas polar cs: radius = 3cm, angle = \x){};
}
\foreach \x/\y in {15/21, 87/22, 159/23, 231/24, 303/25}{
	\node[white_vertex] (\y) at (canvas polar cs: radius = 2.2cm, angle = \x){};
}
\foreach \x/\y in {0/31, 30/32, 72/33, 103/34, 144/35, 174/36,  216/37, 246/38, 288/39, 318/40}{
	\node[white_vertex] (\y) at (canvas polar cs: radius = 1.2cm, angle = \x){};
}
\end{scope}

\draw[solid_edge] (1)--(2)--(3)--(4)--(5)--(6);
\draw[solid_edge] (6)--(7)--(8)--(9)--(10)--(1);
\draw[solid_edge] (1)--(11)--(2);
\draw[solid_edge] (11)--(21);
\draw[solid_edge] (31)--(21)--(32);
\draw[solid_edge] (3)--(12)--(4);
\draw[solid_edge] (12)--(22);
\draw[solid_edge] (33)--(22)--(34);
\draw[solid_edge] (5)--(13)--(6);
\draw[solid_edge] (13)--(23);
\draw[solid_edge] (35)--(23)--(36);
\draw[solid_edge] (7)--(14)--(8);
\draw[solid_edge] (14)--(24);
\draw[solid_edge] (37)--(24)--(38);
\draw[solid_edge] (9)--(15)--(10);
\draw[solid_edge] (15)--(25);
\draw[solid_edge] (39)--(25)--(40);
\draw[solid_edge] (31)--(32)--(35)--(36)--(39)--(40)--(33)--(34)--(37)--(38)--(31);

\end{tikzpicture}
\end{center} \\ [1.4ex] 
    
    Yes & Yes & Yes & All strong snarks & Discussion will follow\tnote{1} \\ 
    \bottomrule
  \end{tabular}
  \caption{Combinations of snark properties that permit equality/inequality in the lower bound on~$c_2(G)$. \vspace{3cm}}
  \label{table:snarkpropertycombos}
  \begin{tablenotes}
  \item [1] Examples and discussion are given in Section~\ref{sec:3conncubic}.
  \end{tablenotes}
  \end{threeparttable}
\end{table}

For each combination of properties except bridgeless, Class~$2$, triangle-free
cubic graphs (i.e. Gardner snarks), we now show that the difference between
the bound and the~$2$-conversion number can be arbitrarily large
(Propositions~\ref{oneconnecteddontmeetbound} to~\ref{trianglesdontmeetbound}%
). We address the remaining category in Section~\ref{sec:3conncubic}, where we
consider~$3$-connected cubic graphs with arbitrary girth.

To prove that the difference between the bound and the~$2$-conversion number
can be arbitrarily large for graphs with bridges, we use the following lemma.

\begin{lemma}
\label{bridgesdonthelp} Let~$G$ be a cubic graph with a bridge~$e$, and let
$H_{1}$ and~$H_{2}$ be the components of~$G-e$. Then~$c_{2}(G) = c_{2}%
(H_{1})+c_{2}(H_{2})$.
\end{lemma}

\noindent\textbf{Proof.\hspace{0.1in}} Clearly,~$c_{2}(G)\leq c_{2}%
(H_{1})+c_{2}(H_{2})$. To show equality we show that  the minimal~$2$-immune 
sets of~$H_{1}$ and~$H_{2}$ (with respect to containment) are the sets $U$ that induce chordless cycles in $G$. 
Let~$U$ be a minimal 
$2$-immune set of~$H_{i}$ and let~$a$ be the vertex of degree~$2$ in~$H_{i}$.
First consider the case where~$a\not \in U$. Then every vertex in~$U$ has
$3$ neighbours in~$H_{i}$ and, since $U$ is $2$-immune, at least $2$ of them are in $U$.  
By the minimality of~$U$, this implies that
$H_{i}[U]$ is a chordless cycle. Now consider the case where~$a\in U$. 
If~$H_{i}[U]$ does not
contain a cycle then it has at least two leaves, and one of these leaves is a
vertex of degree~$3$ in~$H_{i}$. This is a contradiction, since such a vertex
has two neighbours outside~$U$. On the other hand, by
minimality, any cycle in $H_{i}[U]$ contains~$a$
(otherwise the cycle is a smaller~$2$-immune set). 
Therefore, in both cases, the minimal~$2$-immune sets of
$H_{i}$ induce chordless cycles. Since~$G$ is cubic (and therefore its $2$-conversion sets are decycling sets), these are precisely the minimal $2$-immune sets of~$G$. Thus~$U$ is a minimal~$2$-immune
set of~$G$ if and only if it is a minimal~$2$-immune set of~$H_{1}$ or~$H_{2}%
$. Since~$H_{1}$ and~$H_{2}$ are disjoint, the result follows.~$\blacksquare$

\bigskip

\begin{figure}[ht]
\begin{center}
\begin{center}
\begin{tikzpicture}[scale = 0.4,  
white_vertex/.style={whitevertex,inner sep=0pt,minimum size=0.12cm},
black_vertex/.style={blackvertex,inner sep=0pt,minimum size=0.12cm},
solid_edge/.style = {thick, black},
wavy_edge/.style={thick, black, decorate, decoration={snake, segment length = 3mm, amplitude = 0.6mm}},
dotted_edge/.style={thick, black, dashed}]


\begin{scope}
\foreach \x/\y in {0/1, 72/2, 144/3,  216/4, 288/5}{ 
  	\node[white_vertex] (\y) at (canvas polar cs: radius = 1.7cm, angle = \x){};
}
\node[black_vertex] at (2){};
\node[black_vertex] at (5){};

\draw[solid_edge](1)--(2)--(3) --(4) --(5)--(1);
\draw[solid_edge](2)--(4);
\draw[solid_edge](3)--(5);
\node at (canvas polar cs: radius = 3cm, angle = 270){$H_1$};
\end{scope}

\begin{scope}[xshift = 8cm, xscale = 1.2]
\foreach \x/\y in {0/1, 60/2, 120/3,  180/4, 240/5, 300/6}{ 
  	\node[white_vertex] (\y) at (canvas polar cs: radius = 1.7cm, angle = \x){};
}
\node[black_vertex] at (2){};
\node[black_vertex] at (6){};

\draw[solid_edge](1)--(2)--(3) --(4) --(5) --(6) --(1);

\draw[solid_edge](2)--(5);
\draw[solid_edge](3)--(6);

\node at (canvas polar cs: radius = 3cm, angle = 270){$H_2$};
\end{scope}

\begin{scope}[xshift=16cm]
\foreach \y/\x in {1/0, 2/51.4, 3/102.8, 4/154.3, 5/205.7, 6/257.1, 7/308.6}{  
  	\node[white_vertex] (\y) at (canvas polar cs: radius = 2cm, angle = \x){};
}
\node[black_vertex] at (4){};
\node[black_vertex] at (6){};

\draw[solid_edge](1)--(2)--(3) --(4) --(5)--(6)--(7)--(1);
\draw[solid_edge](2)--(5);
\draw[solid_edge](3)--(6);
\draw[solid_edge](4)--(7);

\node at (canvas polar cs: radius = 3cm,  angle =270){$H_3$};

\end{scope}

\begin{scope}[xshift = 24cm, xscale = 1.2]
\foreach \x/\y in {0/1, 60/2, 120/3,  180/4, 240/5, 300/6}{ 
  	\node[white_vertex] (\y) at (canvas polar cs: radius = 1.7cm, angle = \x){};
}
\node[black_vertex] at (2){};
\node[black_vertex] at (5){};

\draw[solid_edge](1)--(2)--(3) --(4) --(5)--(6)--(1);
\draw[solid_edge](2)--(6);
\draw[solid_edge](3)--(5);

\node at (canvas polar cs: radius = 3cm, angle = 270){$H_4$};
\end{scope}

\end{tikzpicture}
\end{center}
\caption{Building blocks for graphs that exceed the bound.}
\label{fig:envelopegraph}
\end{center}
\end{figure}
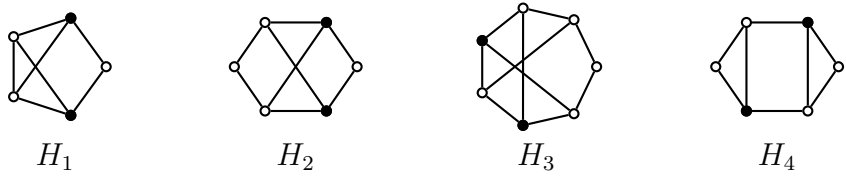

We construct several classes of graphs that exceed the bound from the four
graphs~$H_{1}$,~$H_{2}$, $H_{3}$ and $H_{4}$ shown in Figure
\ref{fig:envelopegraph}.

\begin{lemma}
\label{buildingblocks} Let~$H_{1}$,~$H_{2}$,~$H_{3}$ and~$H_{4}$ be as shown
in Figure~\ref{fig:envelopegraph}, and let~$G$ be a cubic graph containing~$H_{i}$
as an induced subgraph, for some~$1 \le i \le4$. Then any minimum
$2$-conversion set of~$G$ contains exactly~$2$ vertices from each copy of
$H_{i}$.
\end{lemma}

\noindent\textbf{Proof.\hspace{0.1in}} Figure~\ref{fig:envelopegraph} gives a
$2$-conversion set of size~$2$ for each graph~$H_{i}$. On the other hand, no
vertex is on every cycle of~$H_{i}$, so there is no~$2$-conversion set of~$G$
containing fewer than two vertices from any copy of~$H_{i}$.~$\blacksquare$

\bigskip

In Propositions~\ref{oneconnecteddontmeetbound} and
\ref{NoYesYesArbDiff} we construct bridged, Class~$2$ cubic graphs with and
without triangles, respectively, that exceed the bound.

\begin{propn}
\label{oneconnecteddontmeetbound}
Let~$m \ge2$ and let~$G$ be the cubic graph constructed from~$P_{m}$ by
replacing each leaf with a copy of~$H_{1}$ and each internal vertex with a
copy of~$H_{2}$, where~$H_{1}$ and~$H_{2}$ are as shown in Figure
\ref{fig:envelopegraph}. Then

\begin{enumerate}
\item[(a)]~$G$ is a bridged, Class~$2$ cubic graph with triangles, and

\item[(b)]~$c_{2}(G) - \lceil\frac{|V(G)| +2}{4} \rceil= \lfloor\frac{m}%
{2}\rfloor$.
\end{enumerate}
\end{propn}

\noindent\textbf{Proof.\hspace{0.1in}} For (a), the Class~$2$ property follows
from the bridged property by Theorem~\ref{bridgedimpliesclass2}. For (b),
$|V(G)|=6m-2$ and by Lemma~\ref{buildingblocks},~$c_{2}(G)=2m$.~$\blacksquare$

\begin{propn}
\label{NoYesYesArbDiff}
Let~$m \ge2$ and let~$G$ be the cubic graph constructed from~$P_{m}$ by
replacing each leaf with a copy of~$H_{3}$ and each internal vertex with a
copy of~$H_{2}$, where~$H_{2}$ and~$H_{3}$ are as shown in Figure
\ref{fig:envelopegraph}. Then

\begin{enumerate}
\item[(a)]~$G$ is a bridged, Class~$2$, triangle-free cubic graph, and

\item[(b)]~$c_{2}(G) - \lceil\frac{|V(G)|+2}{4}\rceil= \lfloor\frac{m}{2}
\rfloor-1$.
\end{enumerate}
\end{propn}

\noindent\textbf{Proof.\hspace{0.1in}} For (a), the Class~$2$ property follows
from the bridged property by Theorem~\ref{bridgedimpliesclass2}. For (b),
$|V(G)|=6m+2$ and by Lemma~\ref{buildingblocks},~$c_{2}(G)=2m$.~$\blacksquare$

\bigskip

In Proposition~\ref{twoconnecteddontmeetbound} we construct bridgeless, Class
$1$ cubic graphs with and without triangles that exceed the bound.

\begin{propn}
\label{YesNoNoArbDiff}\label{twoconnecteddontmeetbound} Let~$m \ge3$ and let
$H_{2}$ and~$H_{4}$ be as shown in Figure~\ref{fig:envelopegraph}. Let~$G_{1}$
be the cubic graph constructed from~$C_{m}$ by replacing each vertex with a
copy of~$H_{4}$, and let~$G_{2}$ be the cubic graph constructed from~$C_{m}$
by replacing each vertex with a copy of~$H_{2}$. Then

\begin{enumerate}
\item[(a)]~$G_{1}$ is a bridgeless, Class~$1$ cubic graph with triangles,

\item[(b)]~$G_{2}$ is a bridgeless, Class~$1$, triangle-free cubic graph, and

\item[(c)] for~$i=1,2$,~$c_{2}(G_{i}) -\lceil\frac{|V(G_{i})|+2}{4}\rceil=
\lfloor\frac{m-1}{2} \rfloor$.
\end{enumerate}
\end{propn}

\noindent\textbf{Proof.\hspace{0.1in}} Parts (a) and (b) can be easily
verified, using Theorem~\ref{planarimpliesclass1} for (a). For part~(c), it is
clear that~$|V(G_{i})|=6m$ and by Lemma~\ref{buildingblocks},~$c_{2}%
(G_{i})=2m$, for~$i=1,2$.~$\blacksquare$

\bigskip

We have presented cubic graphs with an arbitrary difference between~$c_{2}$
and the lower bound for each of the first four categories defined in Table
\ref{table:snarkpropertycombos}. We now describe a construction that produces
graphs in the fifth category---bridgeless, Class~$2$ cubic graphs of girth~$3$---with an arbitrary difference between~$c_{2}$ and the bound
\eqref{cubiclowerboundceil}. In fact, the same construction can be used to
produce additional examples for any of the girth~$3$ categories.

\begin{figure}[htb]
\begin{center}
\begin{center}
\begin{tikzpicture}[scale = 0.7,  
white_vertex/.style={whitevertex,inner sep=0pt,minimum size=0.12cm},
black_vertex/.style={blackvertex,inner sep=0pt,minimum size=0.12cm},
solid_edge/.style = {thick, black},
wavy_edge/.style={thick, black, decorate, decoration={snake, segment length = 3mm, amplitude = 0.6mm}},
dotted_edge/.style={thick, black, dashed}]


\begin{scope}[rotate = 4]
\foreach \x/\y in {0/1, 30/2, 72/3, 102/4, 144/5, 174/6,  216/7, 246/8, 288/9, 318/10}{ 
  	\node[white_vertex] (\y) at (canvas polar cs: radius = 3.8cm, angle = \x){};
}
\foreach \x/\y in {15/11, 87/12, 159/13, 231/14, 303/15}{   
	\node[white_vertex] (\y) at (canvas polar cs: radius = 3cm, angle = \x){};
}
\foreach \x/\y in {15/21, 87/22, 159/23, 231/24, 303/25}{
	\node[white_vertex] (\y) at (canvas polar cs: radius = 2.2cm, angle = \x){};
}
\foreach \x/\y in {0/31, 30/32, 72/33, 103/34, 144/35, 174/36,  216/37, 246/38, 288/39, 318/40}{
	\node[white_vertex] (\y) at (canvas polar cs: radius = 1.2cm, angle = \x){};
}
\end{scope}

\draw[solid_edge] (1)--(2)--(3)--(4)--(5)--(6);
\draw[solid_edge] (6)--(7)--(8)--(9)--(10)--(1);
\draw[solid_edge] (1)--(11)--(2);
\draw[solid_edge] (11)--(21);
\draw[solid_edge] (31)--(21)--(32);
\draw[solid_edge] (3)--(12)--(4);
\draw[solid_edge] (12)--(22);
\draw[solid_edge] (33)--(22)--(34);
\draw[solid_edge] (5)--(13)--(6);
\draw[solid_edge] (13)--(23);
\draw[solid_edge] (35)--(23)--(36);
\draw[solid_edge] (7)--(14)--(8);
\draw[solid_edge] (14)--(24);
\draw[solid_edge] (37)--(24)--(38);
\draw[solid_edge] (9)--(15)--(10);
\draw[solid_edge] (15)--(25);
\draw[solid_edge] (39)--(25)--(40);
\draw[solid_edge] (31)--(32)--(35)--(36)--(39)--(40)--(33)--(34)--(37)--(38)--(31);

\end{tikzpicture}
\end{center}
\caption{The triangle-replaced graph of the Petersen graph.}
\label{fig:trianglereplacedpetersen}
\end{center}
\end{figure}
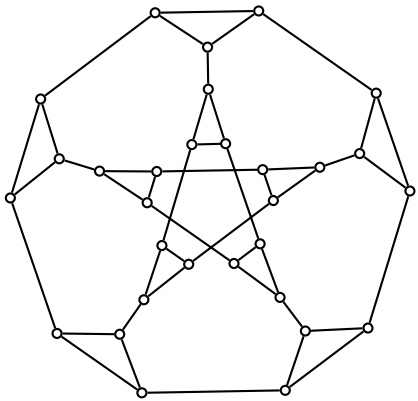

To construct girth~$3$ graphs (which can be bridged or bridgeless and Class
$1$ or Class~$2$) with an arbitrary difference between~$c_{2}$ and the bound
\eqref{cubiclowerboundceil}, we begin with a cubic graph~$G$ and replace each
vertex with a triangle. We call this operation \emph{triangle replacement} of~$G$ 
and we call the resulting girth~$3$ graph the \emph{triangle-replaced
graph} of~$G$, and denote it by~$T(G)$, as in~\cite{WolframTriangleReplaced}.
Lemma~\ref{trianglereplacementstillclass2} guarantees that the
bridged/bridgeless properties and the Class~$1$/Class~$2$ properties are
preserved under triangle replacement. Therefore in order to produce a
bridgeless, Class~$2$ cubic graph with triangles, for example, we take the
triangle replacement of any bridgeless, Class~$2$ cubic graph. Figure
\ref{fig:trianglereplacedpetersen} shows the triangle-replaced graph of the
Petersen graph. Since the Petersen graph is bridgeless and Class~$2$, so is
its triangle-replaced graph.

\bigskip

\begin{lemma}
\label{trianglereplacementstillclass2} For any cubic graph~$G$,~$G$ and~$T(G)$
have the same number of bridges
and the same chromatic index.
\end{lemma}

\noindent\textbf{Proof.\hspace{0.1in}}

The first statement is obvious. For the second statement, let~$G^{\prime
}=T(G)$ and let~$T(v)$ denote the triangle in~$G^{\prime}$ arising from~$v$,
for each vertex~$v$ of~$G$. We consider~$E(G)$ to be a subset of~$E(G^{\prime
})$. We show that~$\chi^{\prime}(G)=3$ if and only if~$\chi^{\prime}%
(G^{\prime})=3$; the result then follows by Vizing's Theorem. Suppose first
that~$G^{\prime}$ has a proper~$3$-edge colouring $f:E(G^{\prime}%
)\rightarrow\{1,2,3\}$. Consider three edges incident with a vertex~$v$ in
$G$. In a proper~$3$-edge colouring of~$G^{\prime}$, these edges all have
different colours, since each is incident with two of the three edges
of~$T(v)$. Therefore the colouring of the edges of~$G$ obtained by restricting~$f$ 
to~$E(G)$ is a proper~$3$-edge colouring of~$G$. Now suppose~$G$ has a
proper~$3$-edge colouring. For each~$v\in V(G)$ we extend the colouring~$f$ to
$T(v)$ such that the edge~$e$ of~$T(v)$ gets the same colour as the edge
of~$E(G)$ that is incident with the other two edges of~$t(v)$.~$\blacksquare$

\bigskip

Lemma~\ref{disjointcycles}, which follows immediately from Proposition
\ref{Prop_DR}(b), gives a lower bound on~$c_{2}(T(G))$, from which we
deduce in Proposition~\ref{trianglesdontmeetbound} that there are
triangle-replaced graphs~$T(G)$ with arbitrary difference between
$c_{2}(T(G))$ and the bound~\eqref{cubiclowerboundceil}.

\begin{lemma}
\label{disjointcycles} Let~$G$ be a~$(k+1)$-regular graph with a collection of
$d$ pairwise disjoint cycles. Then~$c_{k}(G) \ge d$ for all~$k$.
\end{lemma}

We are now ready to show that the difference between the~$2$-conversion number
and the bound~\eqref{cubiclowerboundceil} for triangle-replaced graphs~$T(G)$
grows with the order of~$G$. Since there are arbitrarily large graphs~$G$ for
each feasible category of cubic graphs defined in Table~\ref{table:snarkpropertycombos}, 
there are arbitrarily large differences
between the~$2$-conversion number and the bound for each category with triangles.

\begin{propn}
\label{trianglesdontmeetbound} Let~$H$ be a cubic graph of order~$m$ and let
$G = T(H)$. Then $c_{2}(G) - \lceil\frac{|V(G)|+2}{4} \rceil\ge\lfloor
\frac{m-2}{4} \rfloor$. Moreover,~$G$ has the same number of bridges and the
same chromatic index as~$H$.
\end{propn}

\noindent\textbf{Proof.\hspace{0.1in}} By Lemma~\ref{disjointcycles},
$c_{2}(G)\geq m$. The first statement follows, with~$|V(G)|=3m$. The second
statement follows from Lemma~\ref{trianglereplacementstillclass2}%
.~$\blacksquare$

\bigskip

For each of the first five categories of cubic graphs defined in Table
\ref{table:snarkpropertycombos}, we have given a construction to produce a
graph~$G$ with an arbitrarily large difference between~$c_{2}(G)$ and the
lower bound~$\lceil\frac{|V(G)|+2}{4}\rceil$. However, for all of the
triangle-free graphs, while the difference may be large, the ratio
$\frac{c_{2}(G)}{|V(G)|}$ approaches~$\frac{1}{4}$, and hence the ratio
$\frac{c_{2}(G)}{\lceil\frac{|V(G)+2}{4}\rceil}$ approaches~$1$, as~$|V(G)|$
becomes large. By contrast, for the girth~3 graphs we have constructed in this
section,~$\frac{c_{2}(G)}{|V(G)|}$ approaches~$\frac{1}{3}$ as~$|V(G)|$
becomes large. In the next section we determine whether this ratio can be
greater than~$\frac{1}{4}$, asymptotically, for triangle-free graphs.

\subsection{$3$-edge connected cubic graphs}

\label{sec:3conncubic}\label{here}

In the previous section we constructed infinite families of graphs for which
the difference between the~$2$-conversion number and the lower bound
\eqref{cubiclowerboundceil} could be made arbitrarily large. All of these
examples---in fact, all examples we have seen so far that do not meet the
lower bound---contain triangles or have connectivity at most~$2$. We also saw
infinite families of graphs for which the ratio~$\frac{c_{2}(G)}{|V(G)|}$
exceeds~$\frac{1}{4}$ asymptotically (in~$|V(G)|$), but all of these examples
have girth~$3$. These observations lead us to the following two questions.

\begin{question}
\label{3connarbdiff} Is there a family of~$3$-connected, triangle-free cubic
graphs~$G$ such that~$c_{2}(G) > \left\lceil \frac{|V(G)|+2}{4} \right\rceil
$?

\end{question}

\begin{question}
\label{trifreecubicdontmeetbound} Is there a family of triangle-free cubic
graphs such that
\[
\frac{c_{2}(G)}{|V(G)|} \longrightarrow r >\frac{1}{4} \mbox{ as } |V(G)|
\rightarrow\infty?
\]

\end{question}

In this section we answer both questions in the affirmative. In fact, for
Question~\ref{3connarbdiff} we describe a construction for an infinite family
of~$3$-connected graphs of arbitrary girth such that the difference between
$c_{2}$ and the lower bound~\eqref{cubiclowerboundceil} increases with order.
The same family of graphs provides an answer to Question
\ref{trifreecubicdontmeetbound}.

We begin by defining a graph product that produces an~$r$-regular graph from
two smaller~$r$-regular graphs. In this section we use this product with~$r=3$.

\begin{defn}
\label{GofH} Let~$G$ and~$A$ be~$r$-regular graphs,~$r\geq2$, and define
$A^{-}=A-a$, for any vertex~$a$. Let~$\mathcal{C}$ be the class of graphs that
can be obtained by replacing each vertex~$v$ of~$G$ by a copy~$A_{v}^{-}$ of
$A^{-}$ and joining a degree~$r-1$ vertex of~$A_{u}^{-}$ to a degree~$r-1$
vertex of~$A_{v}^{-}$ if and only if~$uv\in E(G)$. We denote by~$G\circ A^{-}$
any graph in~$\mathcal{C}$.
\end{defn}

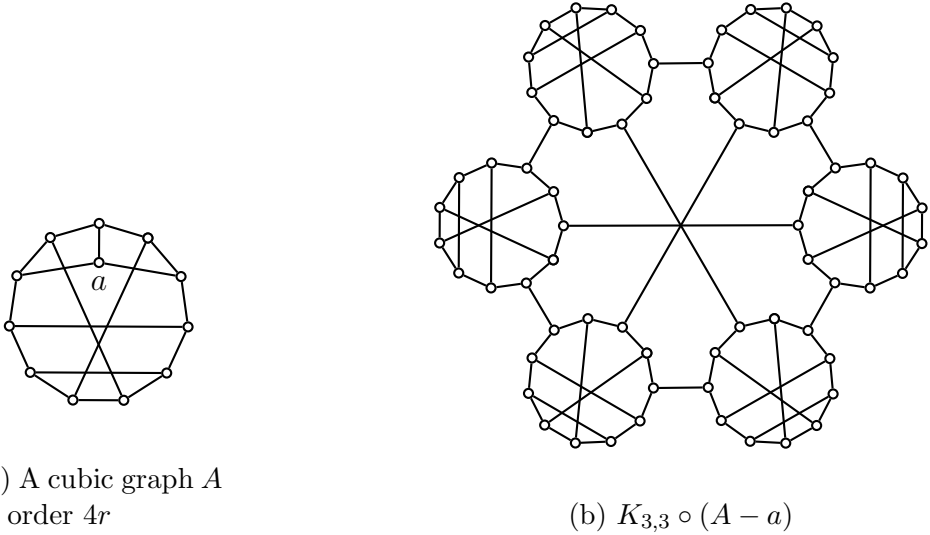
\begin{figure}[hbt]
\centering
\begin{subfigure}[b]{0.2\linewidth}
	\centering
	\begin{center}
\begin{tikzpicture}[scale = 0.4,  
white_vertex/.style={whitevertex,inner sep=0pt,minimum size=0.12cm},
black_vertex/.style={blackvertex,inner sep=0pt,minimum size=0.12cm},
solid_edge/.style = {thick, black},
wavy_edge/.style={thick, black, decorate, decoration={snake, segment length = 3mm, amplitude = 0.6mm}},
dotted_edge/.style={thick, black, dashed}]


\begin{scope}[rotate = 57]
\foreach \x in {0, 1, 2,..., 10,11}{ 
  	\node[white_vertex] (\x) at (canvas polar cs: radius = 3cm, angle = \x*360/11){};
}
\node[white_vertex] (12) at (canvas polar cs: radius = 1.7cm, angle = 360/11){};

\foreach \x [count=\xx from 1] in {0, 1,..., 10}{    
	\draw[solid_edge] (\x)--(\xx);
}

\draw[solid_edge] (2)--(7);
\draw[solid_edge] (4)--(9);
\draw[solid_edge] (5)--(8);
\draw[solid_edge] (6)--(11);
\draw[solid_edge] (1)--(12);
\draw[solid_edge] (3)--(12);
\draw[solid_edge] (10)--(12);
\end{scope}

\node(a label) at ([yshift=-0.7cm]12){$a$};

\end{tikzpicture}
\end{center}\:
	\caption{A cubic graph~$A$ of order~$4r$}
	\label{fig:blobA}
\end{subfigure}\hspace{1cm}
\begin{subfigure}[b]{0.6\linewidth}
	\centering
	\begin{center}
\begin{tikzpicture}[scale = 0.4,  
white_vertex/.style={whitevertex,inner sep=0pt,minimum size=0.12cm},
black_vertex/.style={blackvertex,inner sep=0pt,minimum size=0.12cm},
solid_edge/.style = {thick, black},
wavy_edge/.style={thick, black, decorate, decoration={snake, segment length = 3mm, amplitude = 0.6mm}},
dotted_edge/.style={thick, black, dashed}]

\begin{scope}[rotate = 0, xshift = 6cm, rotate = 147, scale = 0.7]   
	\foreach \x in {0, 1, 2,..., 10,11}{ 
	  	\node[white_vertex] (0\x) at (canvas polar cs: radius = 3cm, angle = \x*360/11){};
	}
	\foreach \x [count=\xx from 1] in {0, 1,..., 10}{    
		\draw[solid_edge] (0\x)--(0\xx);
	}
	\draw[solid_edge] (02)--(07);
	\draw[solid_edge] (04)--(09);
	\draw[solid_edge] (05)--(08);
	\draw[solid_edge] (06)--(011);
\end{scope}

\begin{scope}[rotate = 60, xshift = 6cm, rotate = 147, scale = 0.7]  
	\foreach \x in {0, 1, 2,..., 10,11}{ 
	  	\node[white_vertex] (1\x) at (canvas polar cs: radius = 3cm, angle = \x*360/11){};
	}
	\foreach \x [count=\xx from 1] in {0, 1,..., 10}{    
		\draw[solid_edge] (1\x)--(1\xx);
	}
	\draw[solid_edge] (12)--(17);
	\draw[solid_edge] (14)--(19);
	\draw[solid_edge] (15)--(18);
	\draw[solid_edge] (16)--(111);
\end{scope}

\begin{scope}[rotate = 120, xshift = 6cm, rotate = 147, scale = 0.7] 
	\foreach \x in {0, 1, 2,..., 10,11}{ 
	  	\node[white_vertex] (2\x) at (canvas polar cs: radius = 3cm, angle = \x*360/11){};
	}
	\foreach \x [count=\xx from 1] in {0, 1,..., 10}{    
		\draw[solid_edge] (2\x)--(2\xx);
	}
	\draw[solid_edge] (22)--(27);
	\draw[solid_edge] (24)--(29);
	\draw[solid_edge] (25)--(28);
	\draw[solid_edge] (26)--(211);
\end{scope}

\begin{scope}[rotate = 180, xshift = 6cm, rotate = 147, scale = 0.7] 
	\foreach \x in {0, 1, 2,..., 10,11}{ 
	  	\node[white_vertex] (3\x) at (canvas polar cs: radius = 3cm, angle = \x*360/11){};
	}
	\foreach \x [count=\xx from 1] in {0, 1,..., 10}{    
		\draw[solid_edge] (3\x)--(3\xx);
	}
	\draw[solid_edge] (32)--(37);
	\draw[solid_edge] (34)--(39);
	\draw[solid_edge] (35)--(38);
	\draw[solid_edge] (36)--(311);
\end{scope}

\begin{scope}[rotate = 240, xshift = 6cm, rotate = 147, scale = 0.7] 
	\foreach \x in {0, 1, 2,..., 10,11}{ 
	  	\node[white_vertex] (4\x) at (canvas polar cs: radius = 3cm, angle = \x*360/11){};
	}
	\foreach \x [count=\xx from 1] in {0, 1,..., 10}{    
		\draw[solid_edge] (4\x)--(4\xx);
	}
	\draw[solid_edge] (42)--(47);
	\draw[solid_edge] (44)--(49);
	\draw[solid_edge] (45)--(48);
	\draw[solid_edge] (46)--(411);
\end{scope}

\begin{scope}[rotate = 300, xshift = 6cm, rotate = 147, scale = 0.7] 
	\foreach \x in {0, 1, 2,..., 10,11}{ 
	  	\node[white_vertex] (5\x) at (canvas polar cs: radius = 3cm, angle = \x*360/11){};
	}
	\foreach \x [count=\xx from 1] in {0, 1,..., 10}{    
		\draw[solid_edge] (5\x)--(5\xx);
	}
	\draw[solid_edge] (52)--(57);
	\draw[solid_edge] (54)--(59);
	\draw[solid_edge] (55)--(58);
	\draw[solid_edge] (56)--(511);
\end{scope}

\draw[solid_edge](010)--(13);
\draw[solid_edge](110)--(23);
\draw[solid_edge](210)--(33);
\draw[solid_edge](310)--(43);
\draw[solid_edge](410)--(53);
\draw[solid_edge](510)--(03);
\draw[solid_edge](01)--(31);
\draw[solid_edge](11)--(41);
\draw[solid_edge](21)--(51);

\end{tikzpicture}
\end{center}
	\caption{$K_{3,3} \circ (A - a)$}
	\label{fig:K33circAminus}
\end{subfigure}
\caption{An example of the construction of a cubic graph~$G \circ A^-$.}\label{fig:3conncubicconstruction}
\end{figure}

This construction can yield non-isomorphic graphs depending on~$a$ and on how
the copies of~$A^{-}$ are joined. We will not need to differentiate between
different elements of~$\mathcal{C}$, as our results hold for any such graph.
Figure~\ref{fig:3conncubicconstruction} shows an example of a cubic graph~$A$
with vertex~$a$ identified, and a graph~$K_{3,3}\circ(A-a)$.
.

Proposition~\ref{GcircAminusexceedsbound} asserts that if~$A$ is a cubic graph
of order~$4r$ then~$G\circ A^{-}$ exceeds the bound
\eqref{cubiclowerboundceil}. To answer Question~\ref{3connarbdiff} we then
show that the construction can yield~$3$-edge connected---and therefore
$3$-connected\footnote[2]{The connectivity of any cubic graph is equal to its
edge connectivity~\cite[Theorem 4.6]{ChartrandLesniak}.}--- graphs of
arbitrary girth; this is achieved in Propositions~\ref{GcircAminusgirth} and
\ref{GcircAminus3conn}. We begin with a lemma which guarantees that any
$2$-conversion set of~$G\circ A^{-}$ contains at least~$r$ vertices from each
copy of~$A^{-}$.

\begin{lemma}
\label{atleastrfromAminus} If~$A$ is a cubic graph of order~$4r$ and~$A^{-} =
A - a$ is an induced subgraph of a cubic graph~$H$, then any~$2$-conversion
set of~$H$ contains at least~$r$ vertices of~$A^{-}$.
\end{lemma}

\noindent\textbf{Proof.\hspace{0.1in}} Suppose~$H$ has a~$2$-conversion set
$S$ such that~$|S\cap V(A^{-})|<r$. Then~$(S\cap V(A^{-})\cup\{a\}$ is a
$2$-conversion set of~$A$ of cardinality at most~$r$. However, by
\eqref{cubiclowerboundceil},~$c_{2}(A)\geq\left\lceil \frac{4r+2}%
{4}\right\rceil =\frac{4r+4}{4}=r+1$.~$\blacksquare$

\begin{propn}
\label{GcircAminusexceedsbound} For any cubic graphs~$G$ of order~$n\ge6$ and
$A$ of order~$4r$,
\[
c_{2}(G \circ A^{-}) - \left\lceil \frac{|V(G \circ A^{-})| + 2 }{4}
\right\rceil \ge\left\lfloor \frac{n-2}{4} \right\rfloor .
\]

\end{propn}

\noindent\textbf{Proof.\hspace{0.1in}} Let~$S$ be a~$2$-conversion set of
$G\circ A^{-}$. By Lemma~\ref{atleastrfromAminus},~$S$ contains at least~$r$
vertices of each copy of~$A^{-}$, hence~$|S|\geq nr$. The result follows
because~$V(G\circ A^{-})$ has order~$(4r-1)n$.~$\blacksquare$

\begin{propn}
\label{GcircAminusgirth} Let~$A$ and~$G$ be cubic graphs. Then~$G \circ A^{-}$
has girth at least~$g(A)$.
\end{propn}

\noindent\textbf{Proof.\hspace{0.1in}} Let~$g(A)=g$ and let~$C$ be any cycle
in~$G\circ A^{-}$. If~$C$ is contained in any copy of~$A^{-}$, then~$C$ has
length at least~$g(A)$. If~$C$ is not contained in a copy of~$A^{-}$, then for
any copy~$A_{v}^{-}$ of~$A^{-}$,~$C\cap A_{v}^{-}=\emptyset$ or~$C\cap
A_{v}^{-}$ is a single path, since each copy of~$A^{-}$ is joined by only
three edges to the rest of~$G\circ A^{-}$. Therefore~$C$ consists of segments
$Q_{1},Q_{2},\dots,Q_{s}$ of paths in distinct copies of~$A^{-}$, together
with edges~$e_{i}$ joining~$Q_{i}$ to~$Q_{i+1}$,~$i=1,\dots,s-1$, and~$e_{s}$
joining~$Q_{s}$ to~$Q_{1}$. Each~$Q_{i}$ has length at least~$g-2$, otherwise
$Q_{i}$ and the vertex~$a$ that was removed from~$A$ to form~$A^{-}$ produce a
cycle of length less than~$g$ in~$A$. Therefore~$C$ has length at least
$s(g-2)$. Since~$G$ has no multiple edges,~$s\geq3$, and the result
follows.~$\blacksquare$

\bigskip

We next show that the product~$G\circ A^{-}$ preserves~$3$-connectivity.

\begin{propn}
\label{GcircAminus3conn} Let~$A$ and~$G$ be~$3$-connected cubic graphs. Then
$G \circ A^{-}$ is~$3$-connected.
\end{propn}

\noindent\textbf{Proof.\hspace{0.1in}} Let~$x$ and~$y$ be any distinct
vertices of~$G\circ A^{-}$, say~$x\in V(A_{u}^{-})$ and~$y\in V(A_{v}^{-})$,
for~$u,v\in V(G)$. Let~$u_{i}$ and~$v_{i}$,~$i=1,2,3$, be the vertices of
degree 3 in~$A_{u}^{-}$ and~$A_{v}^{-}$, respectively. 

First, suppose~$u=v$. Since~$A$ is~$3$-connected,~$A$ contains three
internally disjoint~$x-y$ paths, at most one of which contains~$a$. These
correspond to three internally disjoint~$x-y$ paths in~$G\circ A^{-}$: at
least two are contained in~$A_{v}^{-}$ and the third may contain the vertices
$v_{1}$ and~$v_{2}$, say, and a~$v_{1}-v_{2}$ path in~$(G\circ A^{-}%
)-A_{v}^{-}$. 

Now suppose~$u\neq v$. Then in~$A$,~$x$ is connected to~$a$ by three
internally disjoint paths; therefore in~$A^{-}$,~$x$ is connected to the
$u_{i}$'s by three internally disjoint paths. Similarly, in~$A_{v}^{-}$,~$y$
is connected to the~$v_{i}$ by three internally disjoint paths. Since~$G$ is
$3$-connected, there are, without loss of generality, three internally
disjoint paths~$u_{i}-v_{i}$,~$i=1,2,3$. Therefore~$x$ is connected to~$y$ in
$G\circ A^{-}$ by three internally disjoint paths.~$\blacksquare$

\bigskip

Together, Lemma~\ref{GcircAminusexceedsbound} and Propositions
\ref{GcircAminusgirth} and~\ref{GcircAminus3conn} imply that if~$A$ is a
3-connected cubic graph of order~$4r$ and girth~$g$, and~$G$ is a
$3$-connected cubic graph of order~$n\geq6$, then~$G\circ A^{-}$ is a
$3$-connected cubic graph of girth at least~$g$ such that~$c_{2}(G\circ
A^{-})$ exceeds the bound~\eqref{cubiclowerboundceil} by at least
$\left\lfloor \frac{n-2}{4}\right\rfloor~$.
We note that for~$g=3$, we may use~$A=K_{4}$, and then the graph~$G\circ
A^{-}$ is the triangle-replaced graph of~$G$. That is, the~$3$-connected cubic
graphs of girth~$3$ that we presented in Proposition
\ref{trianglesdontmeetbound} are obtainable from the construction presented in
this section.

It remains to show that there exist appropriate cubic graphs~$A$ and~$G$ for
$g\geq4$. For~$G$, we simply require a~$3$-connected cubic graph of order at
least~$6$. There are many such graphs; we highlight one example, which will
also help us find~$A$. 

For~$k\geq2$ and~$g\geq3$, a~$(k,g)$-\emph{cage} is a graph that has the least
number of vertices among all~$k$-regular graphs with girth~$g$. Erd\"{o}s and
Sachs~\cite{ErdosSachs1963}, as cited in~\cite{ChartrandLesniak}, proved that
$(k,g)$-cages exist for all~$k\geq2$ and~$g\geq3$, and Daven and Rodger
\cite{DavenRodger1999} showed that all~$(k,g)$-cages are~$3$-connected.
Therefore a~$(3,g)$-cage is an appropriate choice for~$G$, and if the number
of vertices in such a graph is a multiple of~$4$ then we may use it for~$A$ as
well. (In fact, we may use a~$(3,g_{1})$-cage for~$G$, for any~$g_{1}\geq3$,
and a~$(3,g_{2})$-cage for~$A$, provided that this graph has order~$4r$. The
girth of~$G\circ A^{-}$ will then be at least~$g_{2}$, as shown in Proposition
\ref{GcircAminusgirth}.) 

If, for the specified girth~$g\geq4$, a~$(3,g)$-cage~$B$ has order
$m\equiv2\pmod{4}$, we can obtain a~$3$-connected cubic graph of order~$4r$
and girth at least~$g$ by modifying and joining together two copies of any
$3$-connected cubic graph of order~$4r+2$ and girth at least~$g$ (such as~$B$).

\begin{theorem}
\label{getorder4r}
For every~$g\geq3$ there exists a~$3$-connected cubic graph of order~$4r$ and
girth at least~$g$.
\end{theorem}

\noindent\textbf{Proof.\hspace{0.1in}} For every~$g\geq3$ there exists a
$3$-connected cubic graph with girth~$g$, for example a~$(3,g)$-cage. The
$(3,3)$-cage is~$K_{4}$, so the statement is true for~$g=3$. Let~$g\geq4$ and
suppose~$B$ is a~$3$-connected cubic graph of girth~$g$ and order
$n\equiv2\pmod{4}$. Let~$u$ and~$v$ be two adjacent vertices of~$B$. Since
$g\geq4$,~$u$ and~$v$ have no common neighbour. Let~$a$ and~$b$ be the
neighbours of~$u$ in~$B-v$ and let~$c$ and~$d$ be the neighbours of~$v$ in
$B-u$. Consider two copies~$H$ and~$H^{\prime}$ of~$B-\{u,v\}$; for each
vertex~$v$ in~$H$, we denote its counterpart in~$H^{\prime}$ by~$v^{\prime}$.
Let~$A$ be the cubic graph obtained from~$H$ and~$H^{\prime}$ by adding edges
$aa^{\prime}$,~$bb^{\prime}$,~$cd^{\prime}$ and~$dc^{\prime}$. We show that
$A$ is~$3$-edge connected and has girth at least~$g$. Clearly, any cycle in
$H$ has length at least~$g$, since it is also a cycle in~$B$. Let~$C$ be a
cycle in~$A$ containing vertices from both~$H$ and~$H^{\prime}$ and suppose
$C$ has length~$\ell$. Then, since the vertices~$a^{\prime}$,~$b^{\prime}$,
$c^{\prime}$ and~$d^{\prime}$ are all distinct,~$C\cap H$ is a path~$P$ of
length at most~$\ell-3$ whose endpoints are two of~$a,b,c$ and~$d$. If the
endpoints of~$P$ are~$a$ and~$b$ then~$P+au+ub$ is a cycle in~$B$ of length at
most~$\ell-1$ in~$B$, so~$\ell-1\geq g$. If the endpoints of~$P$ are~$a$ and
$c$, then~$P+au+uv+vc+$ is a cycle in~$B$ of length at most~$\ell$, so
$\ell\geq g$. It remains to show that~$A$ is~$3$-connected. Let~$x$ be any
vertex of~$H$. To see that there are three edge-disjoint~$x-x^{\prime}$ paths
in~$A$, consider three edge-disjoint~$x-v$ paths in~$B$. Without loss of
generality, we may assume that one contains the edge~$au$, another contains
the edge~$cv$ and the third contains the edge~$dv$. Therefore there are paths
$x-a$,~$x-c$ and~$x-d$ in~$H$ and paths~$a^{\prime}-x^{\prime}$,~$c^{\prime
}-x^{\prime}$ and~$d^{\prime}-x^{\prime}$ in~$H^{\prime}$ which are all
edge-disjoint. Adding the edges~$aa^{\prime}$,~$cd^{\prime}$ and~$dc^{\prime}$
produces three edge-disjoint~$x-x^{\prime}$ paths in~$A$. Now let~$x$ and~$y$
be any two vertices of~$H$. Since~$B$ is~$3$-connected,~$H$ is connected.
There are two cases to show that there are three edge-disjoint~$x-y$ paths in
$A$. \smallskip

\noindent\textbf{Case 1:} Suppose there is only one~$x-y$ path~$P$ in~$H$.
Then~$u$ and~$v$ are contained in distinct~$x-y$ paths of~$B$, one of which
contains the subpath~$a-u-b$ and the other contains the subpath~$c-v-d$. Then
$H$ contains edge-disjoint paths~$x-a$,~$b-y$,~$x-c$,~$d-y$, each of which is
disjoint from~$P$, and these paths are copied in~$H^{\prime}$. Therefore~$A$
contains three edge-disjoint~$x-y$ paths,~$(x-a)+aa^{\prime}+(a^{\prime
}-x^{\prime})+(x^{\prime}-c^{\prime})+c^{\prime}d+(d-y)$,~$(x-c)+cd^{\prime
}+(d^{\prime}-y^{\prime})+(y^{\prime}-b^{\prime})+b^{\prime}b+(b-y)$, and~$P$.
\smallskip

\noindent\textbf{Case 2:} Suppose there are exactly two edge-disjoint~$x-y$
paths~$P_{1}$ and~$P_{2}$ in~$H$. Then a third such path in~$B$ contains~$u$
or~$v$ (maybe both), and therefore it contains two of~$a,b,c$ and~$d$, say~$a$
and~$b$ (the other cases are similar). Since~$H^{\prime}$ is connected there
is a path in~$H^{\prime}$ between any two of~$a^{\prime},b^{\prime},c^{\prime
},d^{\prime}$. Then there is a path~$(x-a)+aa^{\prime}+(a^{\prime}-b^{\prime
})+b^{\prime}b+(b^{\prime}-y)$ in~$A$ which is edge-disjoint from~$P_{1}$ and
$P_{2}$. Finally, we must show that for any two vertices~$x$,~$y$ of~$H$,
there are three edge-disjoint~$x-y^{\prime}$ paths in~$A$. Let~$X$ be any
$2$-edge cut in~$A$. Since there are three edge-disjoint~$x-y$ paths in~$A$,
$x$ and~$y$ are in the same component of~$A-X$. Likewise, since there are
three edge-disjoint~$y-y^{\prime}$ paths in~$A$,~$y$ and~$y^{\prime}$ are in
the same component of~$A-X$. Therefore~$x$ and~$y^{\prime}$ are in the same
component of~$A-X$. Since~$X$ is any~$2$-edge cut, there are three
edge-disjoint~$x-y^{\prime}$ paths in~$A$.~$\blacksquare$

\bigskip

We are now ready to answer Question~\ref{3connarbdiff} by proving the
existence of~$3$-connected cubic graphs of arbitrarily large girth that fail
to meet the lower bound~\eqref{cubiclowerboundceil}. However, chromatic index
(either~$3$, corresponding to Class~$1$, or~$4$, corresponding to Class~$2$)
was central to our discussion in the previous section, and we have not yet
discussed the chromatic index of the graphs we have constructed to answer
Question~\ref{3connarbdiff}. In Proposition~\ref{GcircAminuschromaticindex} we
show that the construction produces a Class~$1$ graph if and only if~$G$ and
$A$ are both Class~$1$. We need a lemma, the proof of which can be found in,
e.g.,~\cite[Lemma 4.28]{Wodlinger2018}.

\begin{lemma}
\label{atleast2edgesofeachcolour} If~$H$ is a cubic Class~$2$ graph, then any
$4$-edge colouring of~$H$ contains at least two edges of each colour, and
$H-v$ is Class~$2$ for each~$v\in V(H)$.
\end{lemma}

\begin{propn}
\label{GcircAminuschromaticindex} For any cubic graphs~$G$ and~$A$, the graph
$G\circ A^{-}$ is Class~$1$ if and only if~$G$ and~$A$ are Class~$1$.
\end{propn}

\noindent\textbf{Proof.\hspace{0.1in}} If~$A$ is Class~$2$, then~$A^{-}$ is
Class~$2$, by Lemma~\ref{atleast2edgesofeachcolour}, and therefore~$G\circ
A^{-}$ is Class~$2$. Hence assume~$A$ is Class~$1$. Say~$A^{-}=A-a$ and let
$a_{1},a_{2},a_{3}$ be the vertices of~$A$ adjacent to~$a$. Arguing as in the
proof of Lemma~\ref{atleast2edgesofeachcolour}, we see that in any~$3$-edge
colouring of~$A^{-}$,~$a_{1}$,~$a_{2}$ and~$a_{3}$ are incident with edges
coloured with three different pairs of colours. Assume~$G$ is Class~$1$ and
consider any~$3$-edge colourings of~$G$ and~$A^{-}$ in the same colours.
Colouring the edges~$A_{u}^{-}A_{v}^{-}$ of~$G\circ A^{-}$ the same colour as
$uv$ in~$G$ and suitably permuting the colours in the copies of~$A^{-}$
produces a~$3$-edge colouring of~$G\circ A^{-}$. Now assume~$G$ is Class~$2$
and suppose for a contradiction that~$G\circ A^{-}$ has a~$3$-edge colouring.
For any copy~$A_{v}^{-}$ of~$A^{-}$, let~$xa_{1}$,~$ya_{2}$ and~$za_{3}$ be
the three edges that join~$A_{v}^{-}$ to the rest of~$G\circ A^{-}$. Since
$a_{1}$,~$a_{2}$ and~$a_{3}$ are incident with edges coloured with three
different pairs of colours,~$xa_{1}$,~$ya_{2}$ and~$za_{3}$ have three
different colours. Contracting each copy of~$A^{-}$ to a single vertex yields
$G$ as well as a~$3$-edge colouring of~$G$, which is a
contradiction.~$\blacksquare$

\begin{theorem}
\label{3connarbgirthexceedbound} For any~$g\geq3$ and~$m\in N$, there exists a
$3$-connected cubic graph~$H=G\circ A^{-}$ of girth at least~$g$ such that
$c_{2}(H)-\left\lceil \frac{|V(H)|+2}{4}\right\rceil \geq m$.
Moreover,~$H$ is Class~$1$ if and only if~$G$ and~$A$ are Class~$1$.
\end{theorem}

\noindent\textbf{Proof.\hspace{0.1in}} Theorem~\ref{getorder4r} guarantees the
existence of a~$3$-connected cubic graph of order~$4r$ and girth at least~$g$.
Let~$A$ be such a graph and let~$G$ be any~$3$-connected cubic graph of order
at least~$4m+2$. Then by Propositions~\ref{GcircAminusgirth} and
\ref{GcircAminus3conn},~$H=G\circ A^{-}$ is a~$3$-connected cubic graph of
girth at least~$g$, and by Proposition~\ref{GcircAminusexceedsbound}%
,~$c_{2}(H)$ exceeds the lower bound~\eqref{cubiclowerboundceil} by at
least~$m$. The chromatic index of~$H$ is given by
Proposition~\ref{GcircAminuschromaticindex}.~$\blacksquare$

\bigskip

Any Class 2, girth~$g\geq4$ graph~$G\circ A^{-}$ produced by our construction
is a Gardner snark. For example, taking~$A$ to be the flower snark~$J_{5}$
, a Gardner snark of order~$20$ and
girth~$5$, and any~$3$-connected cubic graph~$G$,~$G\circ A^{-}$ is Class~$2$
(by Proposition~\ref{GcircAminuschromaticindex}),~$3$-connected and has girth
at least~$5$. Therefore it is a Gardner snark (in fact it satisfies a more
restrictive definition of snarks, since it has girth greater than~$4$ and
connectivity greater than~$2$).

We now turn our attention to Question~\ref{trifreecubicdontmeetbound}.
Consider a~$3$-connected cubic graph~$G$ of order~$n$ and a triangle-free
$3$-connected cubic graph~$A$ of order~$4r$, as required for our construction
of the graph~$G\circ A^{-}$. In Lemma~\ref{atleastrfromAminus} we showed that
any minimum~$2$-conversion set of~$G\circ A^{-}$ contains at least~$r$
vertices from each copy of~$A^{-}$. Therefore
\[
\frac{c_{2}(G\circ A^{-})}{|V(G\circ A^{-})|}\geq\frac{rn}{(4r-1)n}=\frac
{r}{4r-1}>\frac{1}{4}.
\]
For example, taking~$A$ to be the graph shown in Figure
\ref{fig:3conncubicconstruction}, and~$G$ any~$3$-connected cubic graph,
$G\circ A^{-}$ has~$\frac{c_{2}(G\circ A^{-})}{|V(G\circ A^{-})|}=\frac{3}%
{11}$.

In fact, it follows from the proof of Lemma~\ref{atleastrfromAminus} that any
$2$-conversion set of~$G \circ A^{-}$ contains at least~$c_{2}(A)-1$ vertices
from every copy of~$A^{-}$, with~$c_{2}(A) \ge r+1$ by
\eqref{cubiclowerboundceil}. Therefore, if~$c_{2}(A) = r +1 + s$,~$s \ge0$,
then every~$2$-conversion set of~$G \circ A^{-}$ contains at least~$r+s$
vertices from each copy of~$A^{-}$. Therefore~$\frac{c_{2}(G\circ A^{-})}{|V(G
\circ A^{-})|} = \frac{r+s}{4r-1}$. That is, by choosing~$A$ to be a cubic
graph of order~$4r$ that does not meet the lower bound
\eqref{cubiclowerboundceil}, we can increase the ratio~$\frac{c_{2}(G\circ
A^{-})}{|V(G \circ A^{-})|}$.

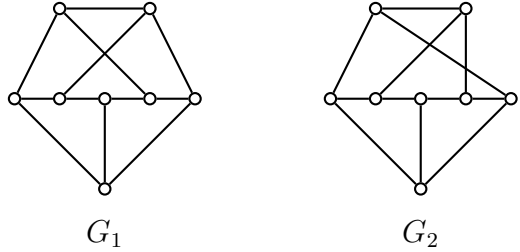
\begin{figure}[htb]
\begin{center}
\begin{tikzpicture}[xscale= 0.6, yscale=0.6,
white_vertex/.style={whitevertex,inner sep=0pt,minimum size=0.15cm},
black_vertex/.style={blackvertex,inner sep=0pt,minimum size=0.15cm},
solid_edge/.style = {thick, black},
wavy_edge/.style={thick, black, decorate, decoration={snake, segment length = 3mm, amplitude = 0.6mm}},
dotted_edge/.style={thick, black, dashed}]

\begin{scope} 
\node[white_vertex](a) at (0,0){};
\node[white_vertex](b) at (1,0){};
\node[white_vertex](c) at (2,0){};
\node[white_vertex](d) at (3,0){};
\node[white_vertex](e) at (4,0){};
\node[white_vertex](f) at (1,2){};
\node[white_vertex](g) at (3,2){};
\node[white_vertex](h) at (2,-2){};

\node at (2, -3) {$G_1$};

\begin{scope}[on background layer] 
	\draw[solid_edge] (a) -- (b) --(c)--(d)--(e);
	\draw[solid_edge] (a) -- (f) --(g) --(e) -- (h) --(a);
	\draw[solid_edge] (h) -- (c);
	\draw[solid_edge] (b) --(g);
	\draw[solid_edge] (d) --(f);
\end{scope}
\end{scope}

\begin{scope}[xshift = 7cm] 
\node[white_vertex](a) at (0,0){};
\node[white_vertex](b) at (1,0){};
\node[white_vertex](c) at (2,0){};
\node[white_vertex](d) at (3,0){};
\node[white_vertex](e) at (4,0){};
\node[white_vertex](f) at (1,2){};
\node[white_vertex](g) at (3,2){};
\node[white_vertex](h) at (2,-2){};

\begin{scope}[on background layer]
	\draw[solid_edge] (h) -- (a) -- (b) --(c)--(d)--(e);
	\draw[solid_edge] (a) -- (f) --(g) --(d);
	\draw[solid_edge] (f) -- (e) -- (h) -- (c);
	\draw[solid_edge] (b) --(g);
\end{scope}
\node at (2, -3) {$G_2$};
\end{scope}

\end{tikzpicture}
\caption{The graphs $G_1$ and $G_2$ of Theorem \ref{cubicupperboundprop}.}
\label{fig:upperboundexceptions}
\end{center}
\end{figure}

Choosing smaller values of~$r$ also increases the ratio. For example, if~$A$
is a cubic graph of order~$8$, then~$c_{2}(A)=3$ (all cubic graphs of order
$8$ meet the lower bound~\eqref{cubiclowerboundceil}) and for any cubic graph
$G$, any~$2$-conversion set of~$G\circ A^{-}$ contains at least two vertices
from each copy of~$A^{-}$. Then~$\frac{c_{2}(G\circ A^{-})}{|V(G\circ A^{-}%
)|}=\frac{2}{7}$. Examples of~$3$-connected cubic graphs of order~$8$ with
girth~$4$---suitable choices for~$A$ in the construction of triangle-free
$3$-connected cubic graphs with ratio~$\frac{2}{7}$---are shown in Figure
\ref{fig:upperboundexceptions}.

For comparison we briefly mention some upper bounds on the~$2$-conversion
number of cubic graphs. Let~$G_{1}$ and~$G_{2}$ be the graphs in Figure
\ref{fig:upperboundexceptions} and let~$\mathcal{G}$ be the class of cubic
graphs obtained from trees, all of whose internal vertices have degree~$3$, by
replacing each internal vertex by a triangle and each leaf by a~$K_{4}$ in
which one edge has been subdivided. 

\begin{theorem}
\label{cubicupperboundprop}\ Let~$G$ be a cubic graph of order~$n>4$.

\begin{enumerate}
\item[(a)] \emph{\cite{Bondy+1987, LiuZhao1996}}\ If~$G\in\mathcal{G}$, then
$c_{2}(G)=\frac{3n+2}{8}$, otherwise~$c_{2}(G)\leq\frac{3n}{8}.$

\item[(b)] \emph{\cite{ZhengLu1990}}\ If~$G$ is triangle-free and
$G\notin\{G_{1},G_{2}\}$, then~$c_{2}(G)\leq\frac{n}{3}.$

\item[(c)] \emph{\cite{Dross+2015}}\ If~$G$ is~$2$-connected, then
$c_{2}(G)\leq\frac{n+2}{3}$ and this bound is sharp.
\end{enumerate}
\end{theorem}

Together, equation~\eqref{cubiclowerboundceil} and Theorem \ref{cubicupperboundprop} bound the value of~$c_{2}(G)$ between~$\left\lceil
\frac{n+2}{4}\right\rceil~$ and~$\left\lfloor \frac{3n+2}{8}\right\rfloor~$
for cubic graphs~$G$ of order~$n>4$. Observe that the ratio~$\frac{c_{2}%
(G)}{|V(G)|}$ cannot exceed~$\frac{1}{3}$ for any triangle-free cubic graph.
It also follows from Theorem ~\ref{cubicupperboundprop} that this ratio is
bounded asymptotically by~$\frac{3}{8}$ for all cubic graphs, and that the
asymptotic bound is attained by the infinite family~$\mathcal{G}$. The graphs
in~$\mathcal{G}$ all have girth~$3$, so the following question remains open.

\begin{question}
\label{quest:largestratio} What is the largest ratio~$\frac{c_{2}(H)}{|V(H)|}$
achievable by an infinite family of~$3$-connected triangle-free cubic graphs
$H$?
\end{question}

\bibliographystyle{plain}
\bibliography{ConversionCubic_Abbr}

%
%
%
%
%
%
%
%
%
%
%
%
%
%
%
%
%
%
%
%
%
%
%
%
%
%
%
%
%
%
%
%
%
%
%
%
%
%

\bigskip
\end{document}